\documentclass[aps,pra, twocolumn,groupedaddress]{revtex4-1} [12pt]
\usepackage[dvips]{graphicx}
\usepackage{bm}
\usepackage{bbm}
\usepackage{color}
\usepackage{mathtools}

\begin{document}

\title{Artificial magnetic fields in momentum space in spin-orbit-coupled systems}

\author{Hannah M. Price$^1$, Tomoki Ozawa$^1$, Nigel R. Cooper$^2$ and Iacopo Carusotto$^1$}
\affiliation{$^{1}$INO-CNR BEC Center and Dipartimento di Fisica, Universit\`{a} di Trento, I-38123 Povo, Italy \\
$^{2}$TCM Group, Cavendish Laboratory, J. J. Thomson Ave., Cambridge CB3 0HE, United Kingdom} 

\bigskip

\bigskip

\begin{abstract}
The Berry curvature is a geometrical property of an energy band which can act as a momentum-space magnetic field in the effective Hamiltonian of a wide-range of systems. We apply the effective Hamiltonian to a spin-1/2 particle in two dimensions with spin-orbit coupling, a Zeeman field and an additional harmonic trap. Depending on the parameter regime, we show how this system can be described in momentum space as either a Fock-Darwin Hamiltonian or a one-dimensional ring pierced by a magnetic flux. With this perspective, we interpret important single-particle properties, and identify analogue magnetic phenomena in momentum space. Finally we discuss the extension of this work to higher spin systems, as well as experimental applications in ultracold atomic gases and photonic systems. 
\end{abstract}

\maketitle

\section{Introduction}

Energy bands with a nontrivial geometry in momentum space are currently of great interest in many fields of physics. The Berry curvature is a geometrical property of an energy band which can be viewed as an artificial magnetic field in a general effective quantum Hamiltonian where the roles of momentum and position are reversed\cite{berry, bliokh2005spin, PhysRevD.12.3845, cooper2012designing, priceozawa, ozawa_momentum_2014}. This has important physical consequences in the anomalous Hall effect\cite{karplus, adams1959energy, nagaosa}, in the collective modes of an ultracold atomic gas\cite{duine, pricemodes} and in the semiclassical dynamics of a wave packet\cite{1chang, dudarev, price, cominotti, dauphin, tomoki}. 

Local geometrical properties, such as the Berry curvature, can also be related to global topological invariants, underlying the quantum Hall effect\cite{thouless} and topological insulators\cite{hasankane, qizhang}. Thanks to recent advances, geometrically nontrivial energy bands have now been created in photonic systems\cite{wang2009, rechtsman2013, hafezi2, jacqmin} and ultracold gases\cite{esslinger,strucksengstock,aidelsburger, hiro}, where the Berry curvature\cite{jotzu, duca} and topological invariants\cite{atala, aidelhof} have been measured directly. 

The consequences of the Berry curvature as a momentum-space magnetic field have been well-studied semiclassically\cite{adams1959energy, 1chang, nagaosa, murakami2003dissipationless, bliokh2005spin, fujita, bliokh2005topological,gosselin2006semiclassical}, but the general effective Hamiltonian can also be exploited as a fully quantum theory\cite{priceozawa}. The effective Hamiltonian describes a single particle in a geometrically nontrivial energy band with an additional external potential. Consequently, analogue magnetic phenomena in momentum space can be explored in the quantum mechanics of single particles in a wide-range of systems; for example, the eigenstates of the Harper-Hofstadter model in an external harmonic trap can be recognised as Landau levels on a torus in momentum space in certain parameter regimes \cite{priceozawa}.

In this paper, we demonstrate that a momentum-space effective magnetic Hamiltonian captures many key features of a single particle in two dimensions with spin-orbit coupling, a Zeeman field, and an additional harmonic trap. spin-orbit-coupled systems are an important area of current theoretical and experimental research \cite{stanescu_spin-orbit_2008, wang2010spin, ho2011bose, zhai2012spin, ozawa_2012, zhou_unconventional_2013, sala, galitski2013spin, li_review_2014, zhaireview}. Recent experiments have realised one-dimensional (1D) spin-orbit coupling in an ultracold gas \cite{solin,zhangdipole}, while various extensions to two dimensions have been proposed\cite{campbell, dalibard,anderson_2013, xuueda}. At the single particle level, the effective momentum-space magnetic Hamiltonian has previously been applied to a system with spin-orbit coupling and a harmonic trap, but without a Zeeman field \cite{cong-jun_unconventional_2011, sinha_trapped_2011, ghosh_trapped_2011, li_two-_2012, ramachandhran_half-quantum_2012, zhou_unconventional_2013}. The addition of a Zeeman field considerably enriches the single-particle phase diagram, revealing a wealth of analogue magnetic phenomena. In this paper, we show, for example, that the effective Hamiltonian can be mapped onto either a Fock-Darwin Hamiltonian or a 1D ring pierced by a tuneable magnetic flux in momentum space depending on the regime of parameters.  In real space, a ring pierced by a magnetic flux supports equilibrium persistent currents even in the ground state; we discuss too how this phenomenon has a direct analogy in momentum space. 

The structure of this paper is the following. First, we introduce the effective momentum-space magnetic Hamiltonian for a general system, then we present the model of a spin-orbit-coupled spin-1/2 particle in two dimensions with a Zeeman field and a harmonic trap. We focus separately on the quantum mechanics of a particle in two parameter regimes; firstly, when the effective Hamiltonian is analogous to the Fock-Darwin Hamiltonian, and secondly, when the momentum-space physics can be understood as that of a 1D ring pierced by a tuneable magnetic flux. We then extend this discussion to higher spin systems, and finally, explore experimental considerations for observing these analogue magnetic effects. 
 
\section{Effective Momentum-Space Magnetic Hamiltonian} \label{sec:effectiveintro}

The effective momentum-space magnetic Hamiltonian can be derived from a generic Hamiltonian: $\mathcal{H} = \mathcal{H}_0 + W ({\bf r})$, where $\mathcal{H}_0$ is either translationally invariant or periodic in real space and $W({\bf r})$, is a weaker additional potential. For the purposes of this paper, $\mathcal{H}_0$ will describe an atom with spin-orbit coupling, with an additional harmonic trap, $W({\bf r}) = \frac{1}{2} \kappa {\bf r}^2$, of trapping strength $\kappa$. 

The eigenfunctions of $\mathcal{H}_0$ for band index $\alpha$ and momentum ${\bf p}$ are $| \chi_{\alpha, {\bf p}} \rangle$. We write these as 
$| \chi_{\alpha, {\bf p}} \rangle = \frac{e^{ i {\bf p} \cdot {\bf r}}}{ \sqrt{V} }| \alpha {\bf p}\rangle$, with $V$ the normalizing volume, to give the Bloch states $| \alpha {\bf p}\rangle$ which have a position dependence that is spatially periodic.
(We set $\hbar=1$ throughout.) For the spin-orbit-coupled systems of this paper, $\mathcal{H}_0$ is translationally invariant. Then the Bloch state, $| \alpha {\bf p}\rangle$, is a spinor, $\eta_\alpha ({\bf p})$, which is independent of position. (The case when $\mathcal{H}_0$ is periodic in real space was previously discussed in Ref.~\onlinecite{priceozawa} and references within.)

The energy bands of $\mathcal{H}_0$ are characterised both by an energy dispersion, $E_\alpha (\mathbf{p})$, and by the geometrical properties of the eigenstates making up the band. In this paper, we focus on two dimensional systems, although the extension to geometrical properties in three dimensions (3D) is straightforward. These geometrical properties are encoded in the Berry connection, $\boldsymbol{\mathcal{A}}_{\alpha} ({\bf p})$, and Berry curvature, $\Omega_{\alpha}({\bf p})$ \cite{berry,di}:
\begin{eqnarray} 
\boldsymbol{\mathcal{ A}}_{\alpha} ({\bf p}) & \equiv & i \langle \alpha {\bf p}| \frac{\partial}{\partial {\bf p}} |\alpha {\bf p}\rangle ,  \label{eq:berrycon}\\
\Omega_{\alpha}({\bf p}) &\equiv & {\bm \nabla_{\bf p}} \times \boldsymbol{\mathcal{ A}}_{\alpha} ({\bf p}) \cdot\hat{{\bf z}}.  \label{eq:berrydef}
\end{eqnarray}  
We note that while the Berry connection is gauge-dependent, the Berry curvature is independent of the gauge choice. 

The eigenfunctions of $\mathcal{H}_0$ are used as a basis in which to expand the eigenstates of the full Hamiltonian, $\mathcal{H}$, as $|\Psi\rangle = \sum_{\alpha}\sum_{\bf p} \psi_{\alpha } (\mathbf{p})  |\chi_{\alpha,\mathbf{p}}\rangle$, where $\psi_{\alpha}(\mathbf{p})$ are expansion coefficients. In general, the additional potential, $W({\bf r})$, mixes different states, $|\chi_{ \alpha {\bf p}}\rangle$. However, provided that the additional potential is sufficiently weak that it does not significantly couple different energy bands, we can assume that the occupation of only band, $\alpha$, is non-negligible. Under this so-called single-band approximation, the effective quantum Hamiltonian is:
\begin{equation}
\tilde{\mathcal{H}}=E_\alpha({\bf p})+W[{\bf r}+ \boldsymbol{\mathcal{A}_\alpha}({\bf p})],
\label{eq:key}
\end{equation}
which is equivalent to a magnetic Hamiltonian with the roles of position and momentum reversed. In particular, the Berry connection, $\boldsymbol{\mathcal{A}}_{\alpha}(\mathbf{p})$, acts like a magnetic vector potential in momentum space, redefining  the relationship between the 
canonical position, ${\bf r}$ and the physical position ${\bf r}+\boldsymbol{\mathcal{A}}_{\alpha} ({\bf p})$ \cite{adams1959energy, nagaosa, bliokh2005spin}. 

To illustrate the duality between real-space and momentum-space magnetism even more clearly, we focus hereafter on the case of an additional harmonic trap $W({\bf r}) = \frac{1}{2} \kappa {\bf r}^2$. Then the effective Hamiltonian can be written as: 
\begin{eqnarray}
	\tilde{\mathcal{H} }	&=& E_\alpha (\mathbf{ p}) + 
	  \frac{\kappa
	\left(i {\bm \nabla}_{\mathbf{p}}
	+
	\boldsymbol{\mathcal{A}}_{\alpha}(\mathbf{p})
	\right)^2}{2}, \label{eq:effective}
\end{eqnarray} 
which is analogous to the textbook magnetic Hamiltonian of a charged particle in an electromagnetic field\cite{Landau1981Quantum}:
\begin{equation}
\mathcal{H}=\frac{(-i {\bm \nabla}_{\mathbf{r}} -e {\bf A}({\bf r}))^2}{2M}+e\Phi({\bf r}) , 
\label{eq:magh}
\end{equation}
where $\Phi({\bf r})$ is a scalar potential, $M$ is the particle mass, $e$ is the particle charge, and ${\bf A}({\bf r})$ is the magnetic vector potential. In these equations, the roles of kinetic and potential energy are reversed, and the harmonic trapping strength, $\kappa$, is equivalent to the inverse particle mass, $M^{-1}$\cite{priceozawa}.

\section{spin-orbit-coupled Spin-1/2 Particle with a Harmonic Trap}

We consider a spin-1/2 particle in two dimensions with Rashba spin-orbit coupling, a Zeeman field and a harmonic trap. The single-particle Hamiltonian is: 
\begin{eqnarray}
\mathcal{H} &=& \mathcal{H}_0 + \frac{1}{2} \kappa {\bf r}^2 \hat{1},  \nonumber \\
\mathcal{H}_0 &=&  \frac{{\bf p}^2}{2M} \hat{1} + \lambda (p_x\hat{\sigma}_y - p_y \hat{\sigma}_x) - \Delta \hat{\sigma}_z  , \label{eq:full}
\end{eqnarray}
where $\hat{\sigma}_{x,y,z}$ are the Pauli matrices, $\hat{1}$ is the 2x2 identity matrix, $\Delta$ is the Zeeman field and $\lambda$ is the spin-orbit coupling strength. Henceforth, we introduce the dimensionless parameters $\zeta  \equiv \lambda^2 M / \Delta$, and $\chi \equiv \frac{ \omega }{\Delta}$ (where $\omega = \sqrt {\kappa / M}$), which compare the spin-orbit coupling and the harmonic trapping strength respectively to the Zeeman field. We note that the limit of a vanishing Zeeman field is captured by $\zeta, \chi \rightarrow \infty$ while $\zeta/\chi$ is kept finite. 

\subsection{The Energy Bands and Eigenstates of $\mathcal{H}_0$}

\begin{figure} [htdp]
\centering
 $
\begin{array}{c}
 (a) \resizebox{0.3\textwidth}{!}{\includegraphics*{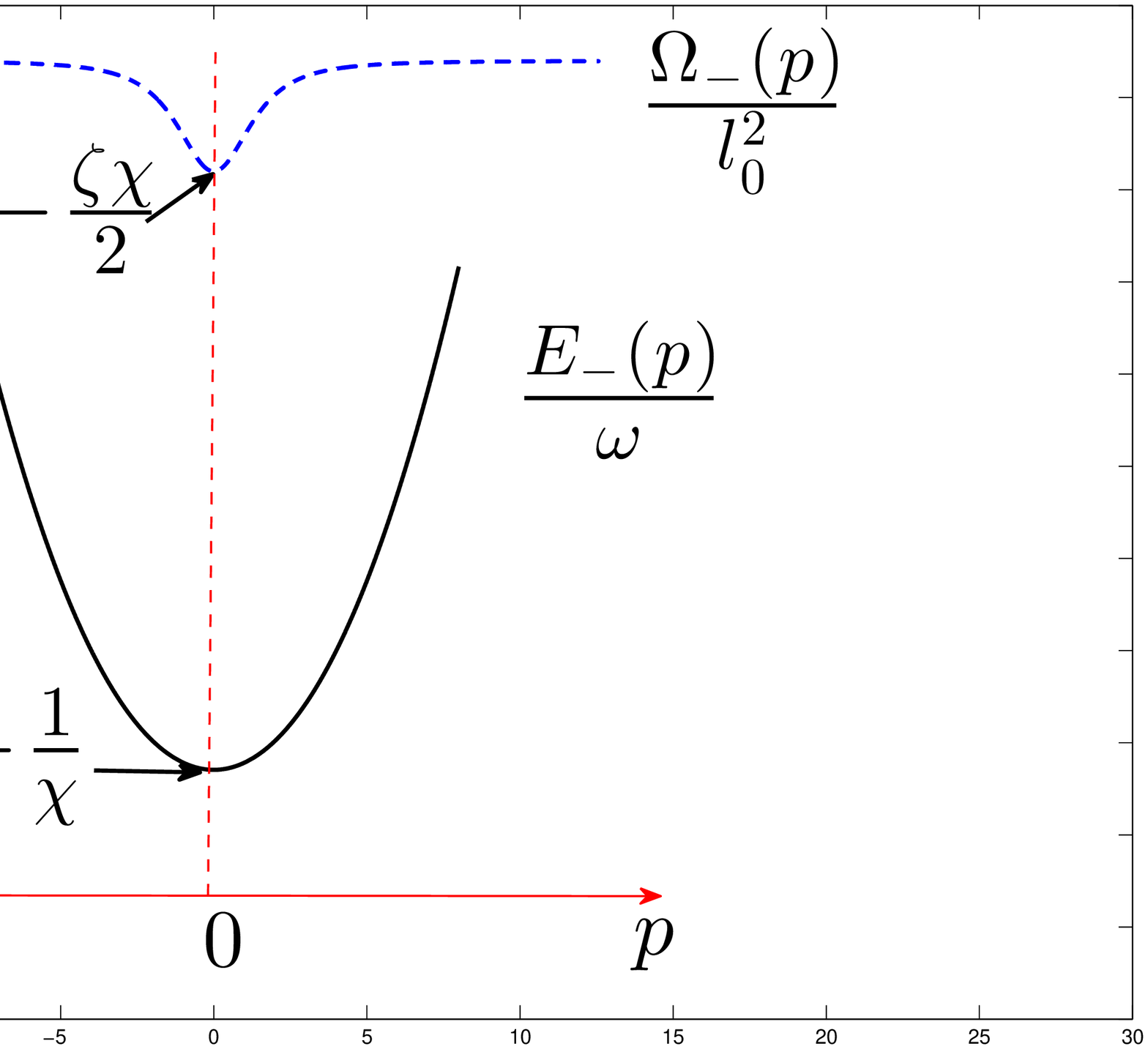}}  \\
  (b) \resizebox{0.35\textwidth}{!}{\includegraphics*{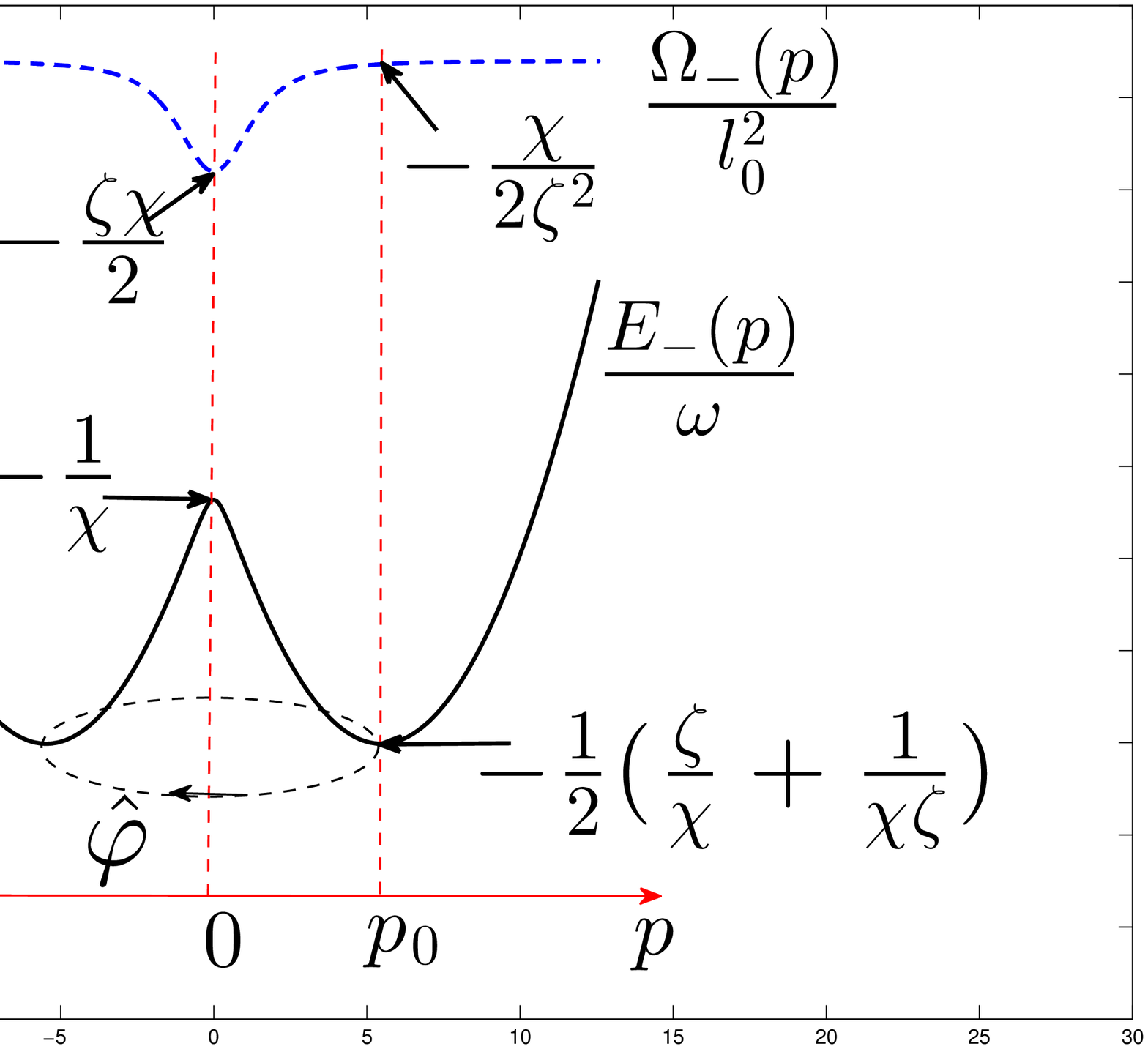}}  
  \end{array}$
\caption{(Color online) A schematic indicating the key characteristics of the energy dispersion and the Berry curvature of the lower band in (a) the single minimum regime for $\zeta = \lambda^2 M / \Delta <1$ and (b) the ring minima regime for $\zeta >1$. } \label{fig:scales}
\end{figure}

Without a harmonic trap, the Hamiltonian reduces to $\mathcal{H}_0$, which has two energy bands\cite{culcer_2003}:
\begin{eqnarray}
E_{\pm} ({\bf p}) = \frac{p^2}{2M}\pm \sqrt{\lambda^2 p^2 + \Delta^2} , \label{eq:dispersion}
\end{eqnarray}
where $\alpha= \pm$ are the band indices. For all parameters, the upper band, $E_+ ({\bf p})$, has a single minimum at ${\bf p}={\bf 0}$. The lower energy band,  $E_- ({\bf p})$, can be tuned between two regimes, illustrated schematically in Fig. \ref{fig:scales}. When the spin-orbit coupling is weak compared to the Zeeman field, $\zeta \equiv \lambda^2 M / \Delta < 1$, the lower band also has a single minimum at ${\bf p}={\bf 0}$. In the opposite limit when $\zeta \equiv \lambda^2 M / \Delta > 1$, the lower band has a ring of minima at $|{\bf p_0}| = \sqrt{\lambda^2 M^2 - \Delta^2 / \lambda^2}$. We refer to these as the single minimum and ring minima regimes, studied in detail in Secs. \ref{sec:single} and \ref{sec:ring} respectively. 

Without a Zeeman field, the lower band is always in the ring minima regime, and the two energy bands are degenerate at ${\bf p}={\bf 0}$. Introducing a Zeeman field lifts the degeneracy and breaks time-reversal symmetry. Then the Berry curvature is non-zero and can be calculated from the spinor eigenfunction of the lower band:
\begin{eqnarray}
\eta_- ({\bf p}) = \sqrt{\frac{1}{2} - \frac{\Delta}{2\sqrt{p^2 \lambda^2 +\Delta^2} }}  
 \left( \begin{array}{c}
i  \frac{\Delta + \sqrt{p^2 \lambda^2 +\Delta^2}}{p \lambda}   \\
e^{  i \varphi} 
  \end{array} \right) \hspace{0.2in} \label{eq:spinor}
\end{eqnarray}
where ${\bf p}= (p, \varphi)$ are the polar coordinates in momentum space. The Berry curvature (\ref{eq:berrydef}) is~\cite{culcer_2003}:
\begin{eqnarray}
\Omega_- ({\bf p}) = -  \frac{\lambda^2 \Delta}{2 (\lambda^2 p^2 + \Delta^2)^{3/2}} .  \label{eq:berry}
\end{eqnarray}
 This is plotted schematically in Figure \ref{fig:scales}. We note that a particle with Dresselhaus spin-orbit coupling, $\lambda (p_x\hat{\sigma}_x + p_y \hat{\sigma}_y)$, instead of Rashba, $\lambda (p_x\hat{\sigma}_y - p_y \hat{\sigma}_x)$, would have the same bandstructure (\ref{eq:dispersion}) but opposite Berry curvature (\ref{eq:berry}). All the following results can be translated from a Rashba to a Dresselhaus spin-orbit-coupled particle by reversing the sign of the Zeeman field, $\Delta$. 

\subsection{The Effective Momentum-Space Hamiltonian}

We now add an external harmonic trap to $\mathcal{H}_0$, and assume that the harmonic trapping strength is sufficiently weak that a single band approximation is valid. This constraint will be discussed quantitatively in the following sections. Following the derivation outlined in Section \ref{sec:effectiveintro}, we combine Eq.~\ref{eq:dispersion} and Eq.~\ref{eq:effective} to write down the effective Hamiltonian for the lower energy band:
\begin{eqnarray}
\tilde{\mathcal{H} }	&=& E_- (\mathbf{ p}) + 
	  \frac{\kappa
	\left(i {\bm \nabla}_{\mathbf{p}}
	+
	\boldsymbol{\mathcal{A}}_{-}(\mathbf{p})
	\right)^2}{2}, \nonumber \\
	&=&  \frac{p^2}{2M} -  \sqrt{\lambda^2 p^2 + \Delta^2}
	+ 
	  \frac{\kappa
	\left(i {\bm \nabla}_{\mathbf{p}}
	+
	\boldsymbol{\mathcal{A}}_{-}(\mathbf{p})
	\right)^2}{2} ,
	 \label{eq:effective2}
\end{eqnarray}
where $\boldsymbol{\mathcal{A}}_{-}(\mathbf{p})$ is the Berry connection (\ref{eq:berrycon}) associated with the Berry curvature of the lower energy band (\ref{eq:berry}). This Hamiltonian will be discussed in more detail in Sections \ref{sec:single} and \ref{sec:ring} for the single minimum and ring minima regimes respectively, where we will show that key properties of this system can be understood in terms of artificial magnetic fields in momentum space.

\subsection{Numerical Calculations based on the \\ full Hamiltonian $\mathcal{H} $} \label{sec:numerics}

\begin{figure} [!]
\centering
 $
\begin{array}{cc}
 \resizebox{0.47\textwidth}{!}{\includegraphics*{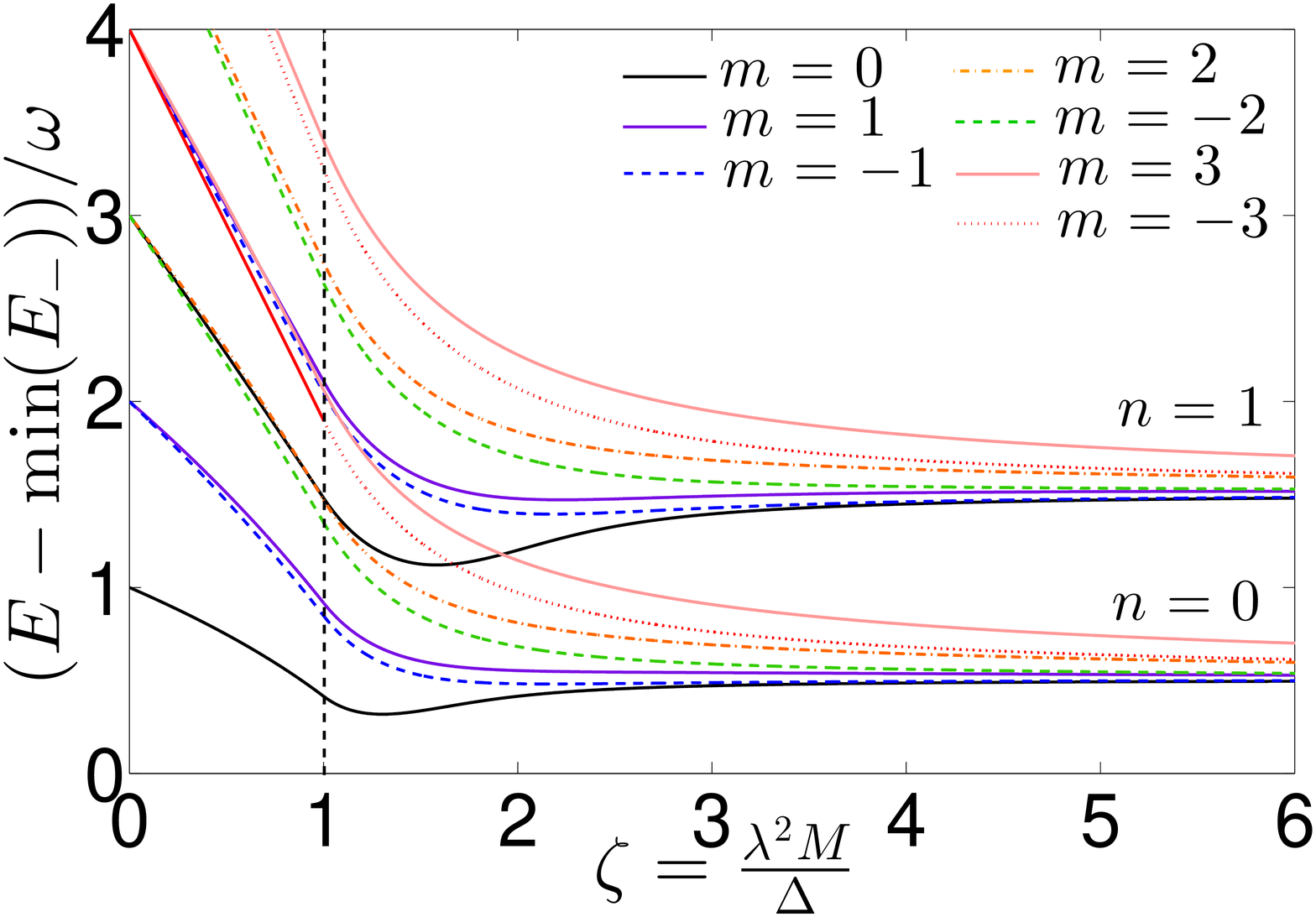}}  &
\end{array} $
\caption{(Color online) The low-energy numerical spectrum adjusted for the minimum energy of the lower band, $\mbox{min}(E_-)$, as a function of $\zeta = \lambda^2 M / \Delta $, for $\chi =0.2$. We have included states with a radial quantum number, $ n=0,1$, and an azimuthal quantum number,  $-3\leq m\leq+3$. The energy of each state is obtained by numerically diagonalizing (\ref{eq:full}) over a basis of 150 states. To emphasise key features, we then subtract the minimum energy of the lower band (\ref{eq:dispersion}) from the total energy of each state. The vertical dotted line marks the cross-over from the single minimum ($\zeta <1$) to the ring minima regime ($\zeta >1$).} \label{fig:overview}
\end{figure}

Throughout this paper, we will compare analytical results from the effective magnetic Hamiltonian in momentum space (\ref{eq:effective2}) with  numerical calculations based on the full Hamiltonian (\ref{eq:full}). To perform these calculations, we expand (\ref{eq:full}) in the basis of single-particle 2D harmonic oscillator states~\cite{ramachandhran_half-quantum_2012} as:
\begin{eqnarray}
\mathcal{H}&=&  \omega ( \hat{a}_d^\dagger \hat{a}_d + \hat{a}_g^\dagger \hat{a}_g + 1 )\hat{1}-  \Delta\hat{\sigma}_z \nonumber \\
&&+ \lambda \sqrt{ M  \omega} \left[ \hat{\sigma}_+ ( \hat{a}_g^\dagger - \hat{a}_d) + \hat{\sigma}_- ( \hat{a}_g - \hat{a}_d^\dagger) \right]  , \label{eq:harm} 
\end{eqnarray}  
where $\hat{\sigma}_\pm = \frac{1}{{2}} ( \hat{\sigma}_x \pm i \hat{\sigma}_y)$ are the spin ladder operators, and where we have introduced the 2D raising and lowering operators: 
\begin{eqnarray} 
\begin{array}{ccc}
\hat{a}_d = \frac{1}{\sqrt{2}} ( \hat{a}_x - i \hat{a}_y)  & \hspace{0.3 in} &\hat{a}_g = \frac{1}{\sqrt{2}} ( \hat{a}_x + i \hat{a}_y)  , 
\end{array} 
\end{eqnarray}
in terms of the standard 1D ladder operators: 
\begin{eqnarray}
\hat{a}_x&=& \sqrt{\frac{M \omega}{2}} \left( \hat{x} + \frac{i}{M\omega} \hat{p}_x \right) ,  
\end{eqnarray}
(and similarly along $y$). Each 2D harmonic oscillator basis state is characterised by two quantum numbers: a radial quantum number, $n=\hat{a}_d^\dagger \hat{a}_d + \hat{a}_g^\dagger \hat{a}_g$, and an azimuthal angular momentum quantum number, $m=\hat{a}_d^\dagger \hat{a}_d - \hat{a}_g^\dagger \hat{a}_g $. 

In this basis, we numerically diagonalise the Hamiltonian (\ref{eq:harm}) to find the eigenspectrum and eigenstates. The effect of the spin-orbit coupling is to couple a ladder of spin-up states with $m$ units of azimuthal angular momentum, to a second ladder of spin-down states with $m+1$ units. Within each ladder, the radial quantum number runs from $n=0,1, ...,\infty$, so that each ladder contains an infinite number of states, equispaced in energy by $2 \hbar \omega$. To perform numerical calculations, we impose an energy cut-off on the basis states, ensuring the cut-off is sufficiently high that all results are converged within the accuracy shown. 

The numerical eigenstates are obtained as a superposition of the 2D harmonic oscillator basis states. To study the spatial properties of an eigenstate, we discretise and superpose the appropriate basis states over a finite lattice of points. We require that the lattice spacing is sufficiently small for results to converge within the accuracy shown. For the calculations in this paper, we chose a discretised grid of 240 by 240 points, with a spacing of $0.02 l_0$, where $l_0 = \sqrt{1/ M \omega}$ is the characteristic real-space lengthscale of the harmonic trap.   

\subsection{Overview of the Numerical Energy Spectrum of the full Hamiltonian $\mathcal{H}$}

Figure \ref{fig:overview} shows the low-energy numerical spectrum of the full Hamiltonian (\ref{eq:harm}) with the minimum energy of the lower band, $\mbox{min}(E_-)$, subtracted. We indicate with a vertical dotted line the cross-over between the single minimum and ring minima regimes, where the functional form of the minimum energy also changes (as shown schematically in Figure \ref{fig:scales}). 

The numerical states can be labelled by the two quantum numbers, $n$ and $m$. By subtracting $\mbox{min}(E_-)$, we emphasise the key features of the numerical spectrum. In particular, deep in the single minimum regime, where $\zeta = \lambda^2 M / \Delta \ll 1$, the principal energy splitting is between the groups of states with different values of $(2n +|m|)$, with a smaller splitting within each group between states with different $m$. In the opposite limit, far into the ring minima regime, $\zeta \gg 1$, the principal energy splitting is between groups of states with different values of $n$, with a smaller splitting within each group between different $m$ states. We shall now show that these hierarchies of energy scales can be understood analytically through the effective momentum-space magnetic Hamiltonian (\ref{eq:effective2}). 

\section{The Single Minimum Regime: Fock-Darwin States}  \label{sec:single}

\subsection{The Effective Momentum-Space Hamiltonian in the Single Minimum Regime} 

When $\zeta \equiv \lambda^2 M / \Delta \ll 1$, we assume that the behaviour of the particle is entirely described by the single-particle states close to the single minimum at ${\bf p}= {\bf 0}$. The band gap at the minimum is $2 \Delta$, and we assume this is much larger than all other energy scales in the system, justifying a single-band approximation. The energy bandstructure is characterised by the effective mass, $M^*$: 
\begin{eqnarray}
E_{-} ({\bf p}) &\simeq& E_0 +  \frac{p^2 }{2 M^*} \\
M^* & =& \frac{1}{(\partial^2 E_- ({\bf p}) / \partial ^2 { \bf p} )} \bigg|_{{\bf p=0}} = \frac{M} {1 -\zeta},  \label{eq:mass}
\end{eqnarray}
where $E_0 = - \Delta$ is the energy of the lower band at  ${\bf p}= {\bf 0}$. The bandstructure is also described by the value of Berry curvature at the minimum (\ref{eq:berry}). For a sufficiently large Zeeman field, $\Delta$, the Berry curvature of the energy band is approximately uniform: 
\begin{eqnarray}
\Omega ({\bf p}) \simeq \Omega_0&=& - \frac{\lambda^2 \Delta}{2 (\lambda^2 p^2 + \Delta^2)^{3/2}} \bigg|_{{\bf p=0}} =- \frac{\lambda^2}{2 \Delta^2}. \label{eq:berrymin}
\end{eqnarray}
The uniform Berry curvature, $\Omega_0$, can be expressed in terms of a Berry connection (\ref{eq:berrycon}). Choosing the symmetric gauge, we express the Berry connection as:
\begin{eqnarray}
\boldsymbol{\mathcal{A}}_- ({\bf p}) = \frac{1}{2}\Omega_0 \left(\begin{array}{c} - p_y\\ p_x\\  0 \end{array} \right) . \label{eq:berryconsingle}
\end{eqnarray}
We substitute Eqs. \ref{eq:mass} \& \ref{eq:berryconsingle} into the effective momentum-space Hamiltonian (\ref{eq:effective2}) to find: 
\begin{eqnarray}
\tilde{\mathcal{H} }	&\simeq& E_0 +  \frac{p^2 }{2 M^*}  +\frac{\kappa}{2} [ - {\bm \nabla}_{\mathbf{p}}^2 + 
i  {\bm \nabla}_{\mathbf{p}} \cdot \boldsymbol{\mathcal{A}}_- ({\bf p})\nonumber\\ && + i\boldsymbol{\mathcal{A}}_- ({\bf p}) \cdot  {\bm \nabla}_{\mathbf{p}}  + (\boldsymbol{\mathcal{A}}_- ({\bf p})) ^2 ] \nonumber \\
&=& E_0 +  \frac{{p}^2 }{2 M^*} - \frac{\kappa}{2} \left( \frac{\partial^2 }{\partial p_x^2} +  \frac{\partial^2 }{\partial p_y^2}  \right)  \nonumber \\&& 
	+ \frac{i \kappa \Omega_0 }{2}\left( - {p}_y \frac{\partial }{\partial p_x} +  {p}_x \frac{\partial }{\partial p_y}  \right)+    \frac{\kappa}{8} \Omega_0^2 {p}^2 . \label{eq:fd}
\end{eqnarray}
Introducing the angular momentum operator: 
\begin{eqnarray}
\hat{L}_z =  (\hat{{\bf r}} \times \hat{{\bf p}})_z = i \left( {p}_y \frac{\partial }{\partial p_x} -  {p}_x \frac{\partial }{\partial p_y}  \right) ,
\end{eqnarray}
we re-write the effective Hamiltonian (\ref{eq:fd}) as: 
\begin{eqnarray}
\tilde{\mathcal{H} }&\simeq&  E_0 - \frac{\kappa}{2} \left( \frac{\partial^2 }{\partial p_x^2} +  \frac{\partial^2 }{\partial p_y^2}  \right)  - \frac{ \kappa \Omega_0 }{2} \hat{L}_{z} \nonumber \\	&&+ \frac{1}{2} \left( \frac{1}{M^*}  +  \frac{\kappa}{4} \Omega_0^2 \right) {p}^2  . \label{eq:12}
\end{eqnarray}
This effective Hamiltonian was derived from a specific model (\ref{eq:full}), but similar results would hold for any system under the single-band approximation where the particle is confined near a minimum in a geometrical energy band where the Berry curvature can be approximated as flat. As we shall now discuss, this Hamiltonian is the momentum-space analogue of the well-known Fock-Darwin magnetic Hamiltonian in real space. 

\subsection{The Fock-Darwin Hamiltonian} 

The Fock-Darwin Hamiltonian describes a charged particle moving in a uniform real-space magnetic field, $ B \hat{{\bf z}}$, and a harmonic potential. This theoretical model was first proposed over eighty years ago, independently by both Fock\cite{fock_1928} and Darwin\cite{darwin1931diamagnetism}. Since then, despite its simplicity, the Fock-Darwin Hamiltonian has found important practical applications, for example, in describing few-electron quantum dots under relatively weak magnetic fields\cite{mceuen_1992, ashoori_1993, schmidt1995quantum,  tarucha_1996}. In these systems, the additional energy required to add one more electron to a quantum dot can be modelled, in certain regimes, as that of adding a non-interacting particle, described by the Fock-Darwin Hamiltonian, plus a constant interaction energy\cite{chakraborty1999quantum, kouwenhoven2001few, reimann_2002}. 

The Fock-Darwin Hamiltonian follows straightforwardly from the general magnetic Hamiltonian (\ref{eq:magh}) in real space with an additional harmonic potential: 
\begin{eqnarray}
\mathcal{H}_{FD} &=&  \frac{({{\bf p}} - e {\bf A})^2}{2M} + \frac{1}{2} \kappa' {\bf r}^2 \nonumber \\
&=&
 - \frac{1}{2 M} \left( \frac{\partial^2 }{\partial x^2} +  \frac{\partial^2 }{\partial y^2}  \right)  + \frac{ e B }{2 M } \hat{L}_{z} \nonumber \\	&&+  \frac{1}{2}   \left(  \kappa' +  \frac{e^2 B^2 }{4 M}  \right) \hat{r}^2  , \label{eq:fdreal}
\end{eqnarray}
where $\kappa' $ is the strength of a harmonic trap with frequency, $\omega'$, and we have chosen the symmetric gauge for the magnetic vector potential:
\begin{eqnarray}
{\bf A} ({\bf r})=  \frac{1}{2}B  \left(\begin{array}{c} - y\\ x\\  0 \end{array} \right) . 
\end{eqnarray}
Comparing Eqs.  \ref{eq:12} and \ref{eq:fdreal}, we see that the two Hamiltonians have the same form (up to minus signs and an energy shift, $E_0$). Translating between these Hamiltonians, the roles of position and momentum are reversed. In particular, the particle mass, $M$, and the harmonic trapping strength, $\kappa'$ in real space are replaced by, respectively, the inverse harmonic trapping strength, $\kappa^{-1}$ and the inverse effective mass, $(M^*)^{-1}$ in momentum space. The Berry connection, $\boldsymbol{\mathcal{A}}_- ({\bf p})$ is analogous to the magnetic vector potential, $e {\bf A} ({\bf r})$, while the Berry curvature, $\Omega_0$, plays the role of the uniform magnetic field, $eB$. 

\subsection{The Fock-Darwin Eigenstates and Eigenspectrum}

We use the duality between Eqs. \ref{eq:12} \& \ref{eq:fdreal}, to translate the known energy spectrum and eigenstates of the Fock-Darwin Hamiltonian from real space into momentum space. 

Ignoring the angular momentum term, the Fock-Darwin Hamiltonian (\ref{eq:fdreal}) is that of a 2D harmonic oscillator with a shifted frequency:
\begin{eqnarray}
\omega_F = \sqrt{ \frac{\kappa'}{M}+  \frac{1}{4} \omega_c^2} ,
\end{eqnarray}
where $\omega_c= e B / M$ is the cyclotron frequency. As the 2D harmonic oscillator eigenstates are also eigenstates of $\hat{L}_z$, the full Fock-Darwin eigenspectrum follows directly as \cite{fock_1928, darwin1931diamagnetism}:
\begin{eqnarray}
E_{n,m}' &=& (2n+1+|m|) \omega_F + \frac{1}{2} m \omega_c ,  \label{eq:fdspec}
\end{eqnarray}
where $n$ is the radial quantum number, and $m$ is the azimuthal quantum number. From the analogy between the real-space Fock-Darwin Hamiltonian and the effective energy band Hamiltonian, we translate this energy spectrum into momentum space:  
\begin{eqnarray}
E_{n,m}&=& (2n+1+|m|) \sqrt{\frac{\kappa}{M^*}+\frac{\kappa^2 \Omega_0^2}{4}} - \frac{1}{2} m \kappa \Omega_0  + E_0, \nonumber \label{eq:enmcin}
\end{eqnarray}
where $\kappa \Omega_0$ is the analogue cyclotron frequency. The energy spectrum is:  
\begin{eqnarray}
\frac{E_{n,m} }{\omega}&=&  (2n+1+|m|) \sqrt{ ( 1 - \zeta) + \frac{(\zeta\chi)^2}{16}} \nonumber \\
&&+ \frac{1}{4}  m \chi \zeta - \frac{1}{\chi} ,   \label{eq:enfd}
\end{eqnarray}
in terms of the dimensionless parameters defined above.

\begin{figure} [htdp]
\centering
 $
\begin{array}{c}
(a) \resizebox{0.45\textwidth}{!}{\includegraphics*{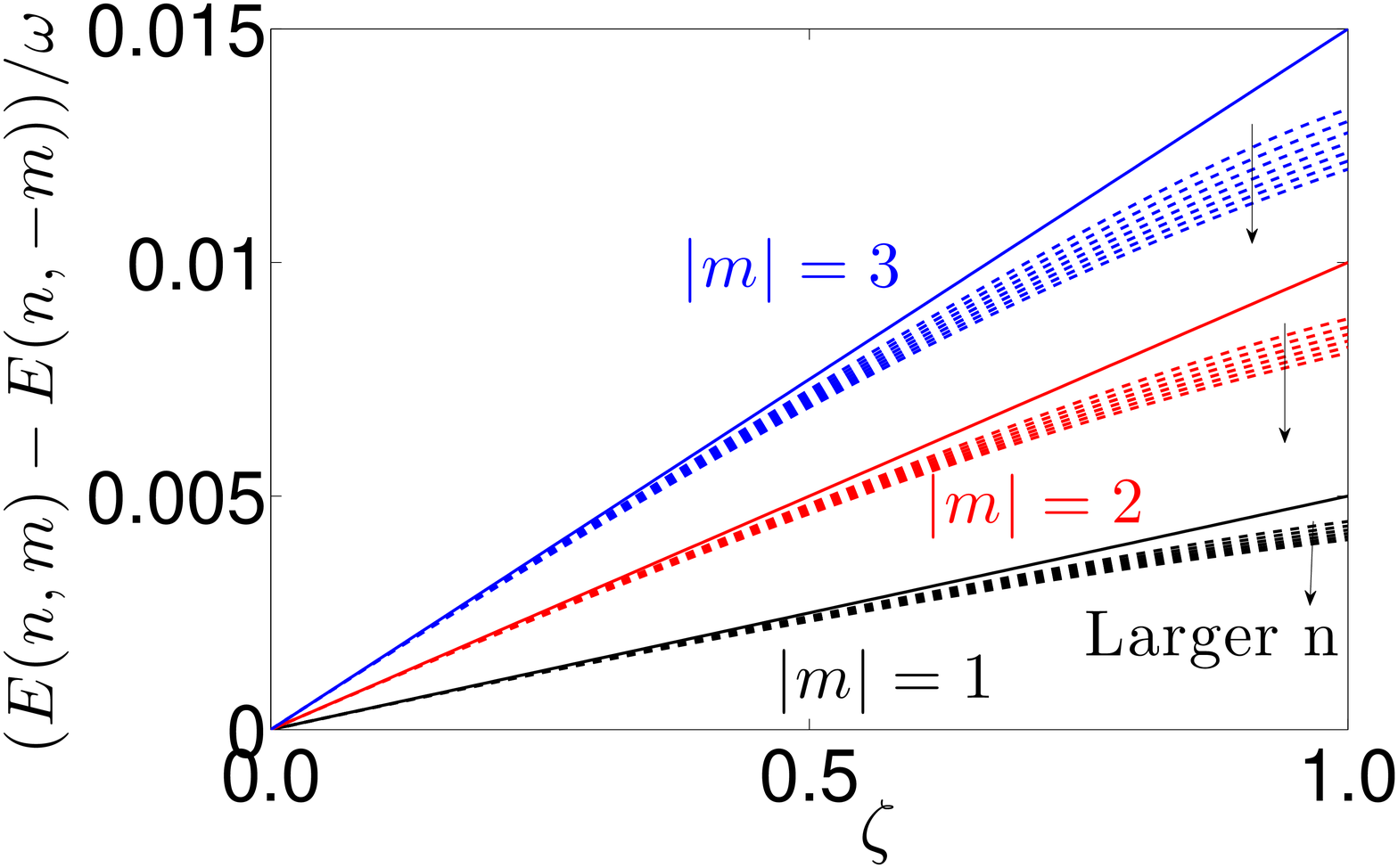}}  \\
(b) \resizebox{0.45\textwidth}{!}{\includegraphics*{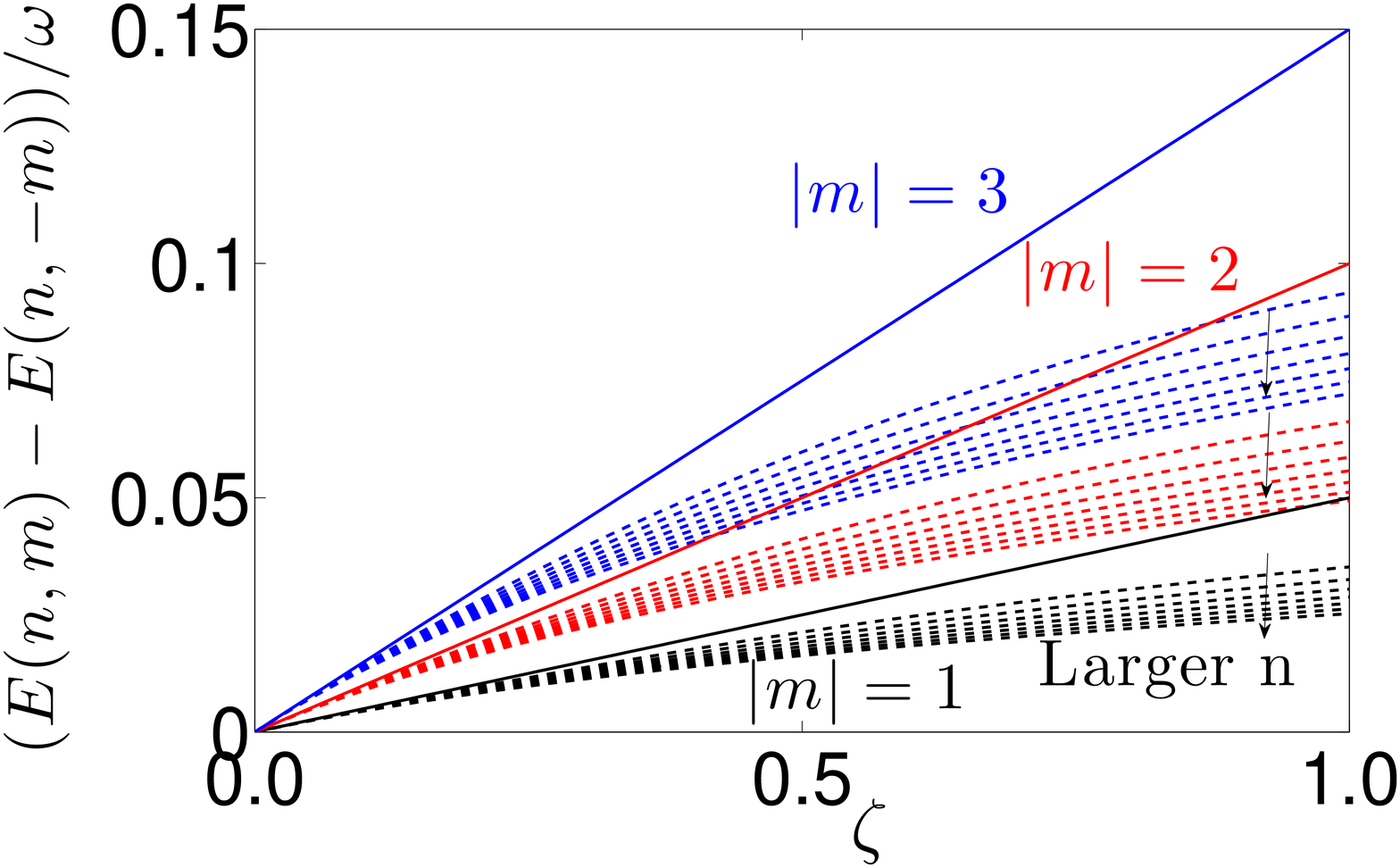}}  
\end{array} $
\caption{(Color online) The energy splitting between states with $+m$ and $-m$ for (a) $\chi=0.01$ and (b)  $\chi=0.1$. {\it Solid Lines} are the analytical Fock-Darwin results (Eq. \ref{eq:split}). {\it Dashed Lines} are numerical calculations from the full Hamiltonian (\ref{eq:full}). Multiple dashed lines of the same color represent states with different values of $n$. The agreement between analytics and numerics is best for small $\chi$, $\zeta$, $n$ and $m$.} \label{fig:msplit}
\end{figure}

Similarly, we can write down the Fock-Darwin states analytically, as these are 2D harmonic oscillator states in real space, with a modified characteristic length scale, $l_F = \sqrt{{\kappa'/  \omega_F}}$\cite{chakraborty1999quantum, kouwenhoven2001few}. Translating between real space and momentum space, the analytical form of the eigenstates in momentum space is:
\begin{eqnarray}
\psi_{n, m} ( p, \varphi)  &=& \sqrt{\frac{n!}{\pi (n + |m|)!}} 
 \frac{p^{|m|}}{ l_{\Omega}^{ (|m| + 1)}} e^{ i m \varphi} \nonumber \\ &&\times e^{- p^2/2 l_\Omega^2} L_n^{|m|} ( p^2 / l_\Omega^2) , \label{eq:eigen}
\end{eqnarray}
where $L_n^{|m|}$ are the generalised Laguerre polynomials and: 
\begin{eqnarray}
l_\Omega
&=& \frac{1}{l_0} \left( ( 1 - \zeta) + \frac{(\zeta\chi)^2}{16}\right)^{-1/4}  , \label{eq:length}
\end{eqnarray}
is the characteristic momentum scale, analogous to $l_F$. As $\zeta \rightarrow 0$, the characteristic momentum scale reduces to the inverse simple harmonic oscillator length: $l_\Omega \rightarrow 1/l_0$. This corresponds to the limit in which the effective Hamiltonian (\ref{eq:12}) is that of a simple harmonic oscillator in momentum space with no artificial magnetic field. 

These analogue Fock-Darwin states (\ref{eq:eigen}) are characterised in momentum space by the qualitative features of 2D harmonic oscillator states in real space. The quantum number, $n$, counts the number of radial nodes lying away from the origin, while a non-zero angular momentum, $|m|$, introduces an additional node at ${\bf p} = {\bf 0}$. States with $+m$ and $-m$ are the same up to a phase factor $e^{ \pm i m \varphi}$, while increasing values of $|m|$ lead to wider dips in the density around the origin \cite{kouwenhoven2001few}. 

\subsection{Comparison of Analytical and \\Numerical Results} 

We now compare the analytical Fock-Darwin eigenspectrum and eigenstates in momentum space, presented above, with the numerical calculations outlined in Section \ref{sec:numerics}, based on the full Hamiltonian (\ref{eq:full}). 

In the analytical eigenspectrum (\ref{eq:enmcin}), the energy splitting between different values of $(2n+|m|)$ is larger than the splitting between different $m$ as $\zeta<1$ in the single minimum regime. This separation of scales is displayed in the numerical energy spectrum in Figure \ref{fig:overview}. As can be seen from Eq. \ref{eq:enmcin}, the magnitude of these energy scales is controlled by $\chi$, which measures the strength of the harmonic trap relative to the Zeeman field. In the effective Hamiltonian, we have used the single-band approximation, which assumes that the harmonic trap does not significantly couple the two energy bands. The minimum band gap is $2\Delta$ at ${\bf p=0}$, and so we take $\chi \ll 2$ to ensure our analytical interpretation is valid. 

In the limit that $\zeta \rightarrow 0$, the Berry curvature vanishes and the energy spectrum is that of a simple 2D harmonic oscillator in the absence of a magnetic field: 
\begin{eqnarray}
\frac{E_{n,m} - \mbox{min}(E_-) }{\omega}\rightarrow (2n+1+|m|)  , 
\end{eqnarray}
where $\mbox{min}(E_-) = E_0$ is the energy offset from the minimum energy of the lower band. When the Berry curvature is zero, the states with the same value of $(2n+|m|)$ are degenerate. This behaviour can be clearly seen in the $\zeta \rightarrow 0$ limit of the numerical low energy spectrum in Figure \ref{fig:overview}. 

As $\zeta$ increases, the momentum-space magnetic field splits the degenerate eigenstates. In the simplest case, for two eigenstates with the same value of $n$ and $\pm m$, the energy splitting is given by: 
\begin{eqnarray}
\frac{E_{n,m} - E_{n,-m} }{\omega} =  \frac{1}{2}  m \chi \zeta . \label{eq:split}
\end{eqnarray}
The splitting of states with $n=0, m= \pm 1$ could be experimentally measured in the splitting of the dipole mode frequency of an ultracold gas\cite{duine, pricemodes}. This is because, for the special case of the dipole mode, interactions drop out and the dipole mode of this system can be understood at this single-particle level\cite{pricemodes}. 

The analytical prediction (\ref{eq:split}) is compared with numerical results in Fig. \ref{fig:msplit} for a range of parameters. As can be seen, the agreement with numerics is excellent for small values of parameters $\chi$, $\zeta$, $n$ and $m$, but breaks down as these parameters are increased. This is because we have assumed that the system is well described by an effective mass and a uniform Berry curvature, both defined at the minimum ${\bf p}={\bf 0}$. In fact, there will be higher order terms which are not captured by our analytical Fock-Darwin spectrum (\ref{eq:enmcin}). We expand both the effective mass approximation and the Berry curvature to next highest order, assuming that the eigenstates vary on the characteristic momentum scale, $\delta p \propto l_\Omega$. The next-order corrections scale as: 
\begin{eqnarray}
\frac{\delta E_-}{\omega} &=& \frac{1}{\omega} \frac{1}{4!} \frac{\partial ^4 E_-({\bf p})}{\partial {\bf p}^4} \bigg|_{{\bf p}=0}  (\delta p)^4 \propto  \frac{ \zeta^2 \chi}{ ( 1 - \zeta) + \frac{(\zeta\chi)^2}{16}}  \nonumber \\
\delta \Omega_- &=& \frac{1}{2} \frac{\partial ^2 \Omega_- ( {\bf p})}{\partial {\bf p}^2}  \bigg|_{{\bf p}=0} ( \delta p)^2  \propto  \frac{\chi \zeta} {{\sqrt{ ( 1 - \zeta) + \frac{(\zeta\chi)^2}{16}}}}   |\Omega_0| , \nonumber  \label{eq:qucorr}
\end{eqnarray}
which decrease as $\zeta, \chi \rightarrow 0$. The corrections will also be smaller for lower values of $n$ and $|m|$, where the states are more localised in momentum space. Our analytical interpretation therefore best describes numerics when the parameters are small. 
We note that $\delta \Omega_-$ has the opposite sign to $\Omega_0$ and therefore will reduce the effective momentum-space magnetic field and the splitting between states. This is observed in Fig. \ref{fig:msplit}, where the numerical results universally lie below the analytical prediction. 

In the limit $\zeta \rightarrow 1$, the discrepancy between analytics and numerics can be large, as shown in Fig. \ref{fig:msplit}. This is because there is a transition in this limit between the single minimum and ring minima regimes where the effective mass, $M^*\rightarrow \infty$. The divergence of the effective mass is analogous to a vanishing harmonic potential in the real-space Fock-Darwin Hamiltonian (\ref{eq:fdreal}), where the Fock-Darwin states tend towards Landau levels. However, as seen from the corrections discussed above, the energy dispersion also provides an effective quartic trapping in momentum space, which does not vanish as $\zeta \rightarrow 1$. 

\begin{figure} [htdp]
\centering
 $
\begin{array}{c}
(a) \resizebox{0.45\textwidth}{!}{\includegraphics*{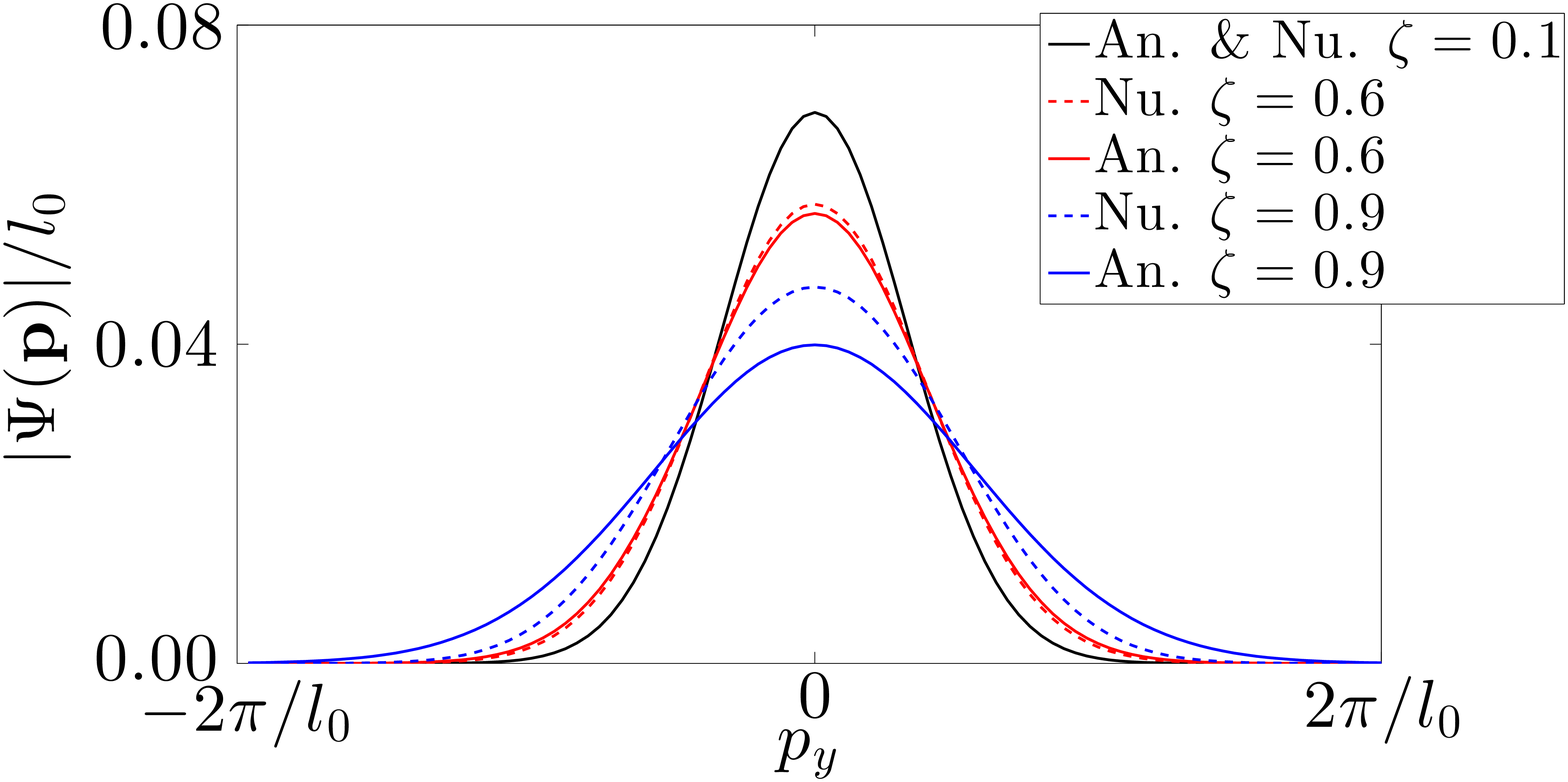}}  \\
(b) \resizebox{0.45\textwidth}{!}{\includegraphics*{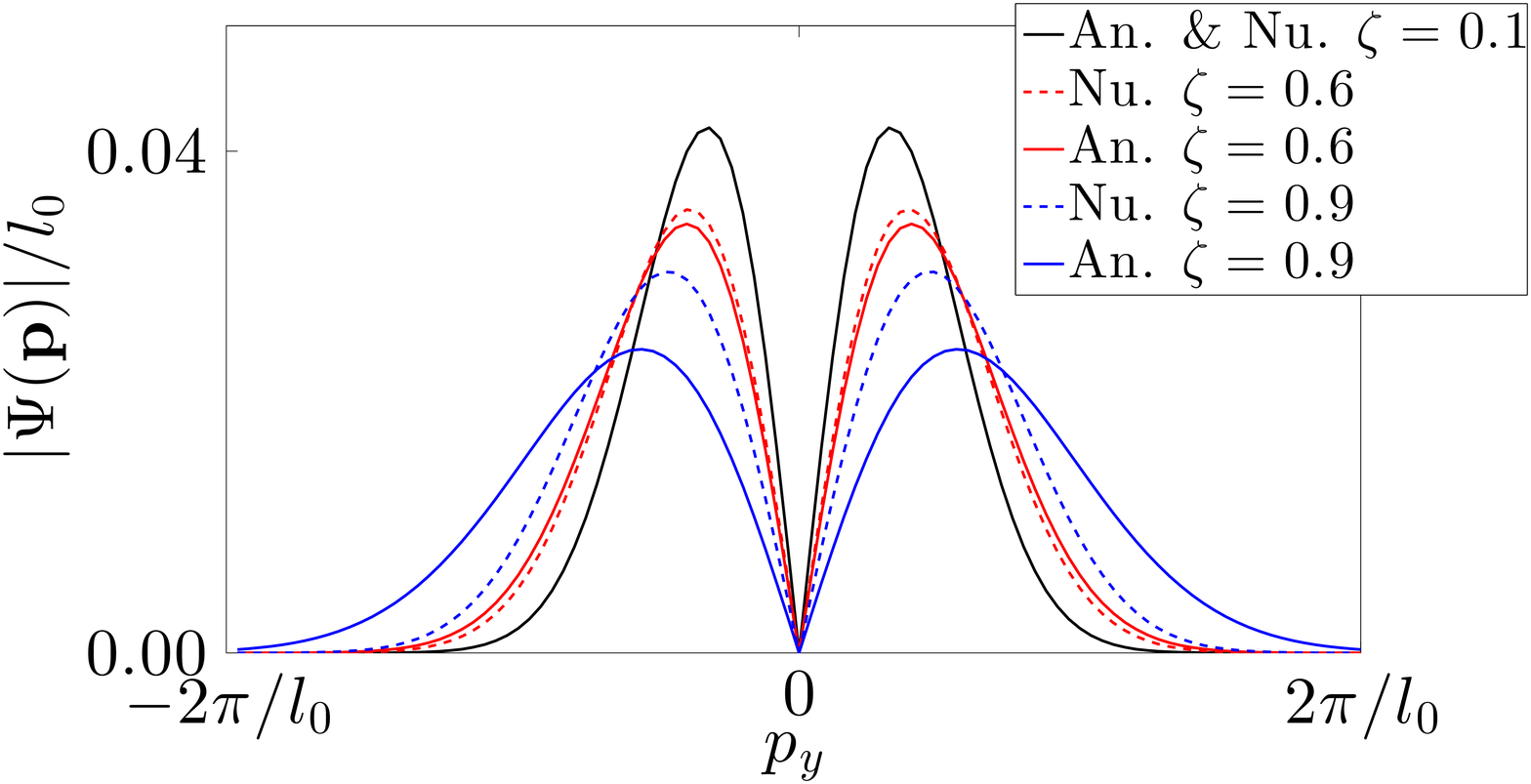}}  
\end{array} $
\caption{(Color online) Quantitative comparison between the analytical (An.) Fock-Darwin eigenstates (\ref{eq:eigen}) and numerical (Nu.) eigenstates of the full Hamiltonian (\ref{eq:full}) for (a) $n=0$, $m=0$ and (b) $n=0$, $m=1$ for $\chi=0.1$ and over a range of $\zeta$.} \label{fig:quant}
\end{figure}

The above observations are supported by a comparison of the analytical wave function (\ref{eq:eigen}) with the numerically calculated eigenstate of the full Hamiltonian (\ref{eq:full}) along $p_x=0$ for different values of $\zeta$, shown in Figure \ref{fig:quant}. As can be seen, the agreement again improves for lower energy states and as $\zeta$ decreases. In particular, as $\zeta \rightarrow 1$, the additional quartic potential from the energy band dispersion becomes increasingly important. Thanks to this additional trapping potential, the wave functions are more tightly confined in momentum space than expected from the simple Fock-Darwin description. This is illustrated in Fig. \ref{fig:quant}, where for small $\zeta=0.1$, the theoretical and numerical results are indistinguishable. For large $\zeta=0.9$, conversely, theory and numerics quantitatively disagree, and the theoretical prediction over-estimates the spread of the wave function in momentum space. 

\section{Ring Minima Regime}  \label{sec:ring}

\subsection{The Effective Momentum-Space Hamiltonian in the Ring Minima Regime}

For strong spin-orbit coupling or a weak Zeeman field, $\zeta \equiv \lambda^2 M / \Delta > 1$ and the lower energy band of $\mathcal{H}_0$ has a ring of degenerate minima. The effective momentum-space Hamiltonian in this regime has previously been studied for the special case of a vanishing Zeeman field $\Delta =0$ \cite{cong-jun_unconventional_2011, sinha_trapped_2011, ghosh_trapped_2011, li_two-_2012, ramachandhran_half-quantum_2012, zhou_unconventional_2013}. We now generalise the effective Hamiltonian to include a Zeeman field, revealing a rich phenomenology. 

We assume that the trap is much weaker than the energy difference $ \Delta E = [E_-({ p_0})-E_-({ 0})] $ (Figure \ref{fig:scales}), allowing us to assume that, at low energies, a particle is confined to the ring of minima. Under this constraint, the single-band approximation is automatically justified, as the energy of the lower band at the origin, $E_-({\bf 0})$, is always less than or equal to the minimum energy of the upper band, $E_+ ({\bf p})$. When the particle is confined around the ring, we can make a separable ansatz for the wave function\cite{zhou_unconventional_2013, chen_2014}:
\begin{eqnarray}
\psi_{nm} ({\bf p}) \simeq G_m (\varphi ) \frac{f_n(p)}{\sqrt{p} } ,  \label{eq:eigenstate}
\end{eqnarray}
where $G_m (\varphi )$ obeys an angular and $f_n(p)$ a radial effective Hamiltonian (and the functions are indexed by $m$ the azimuthal quantum number and $n$ the radial quantum number respectively). We now discuss these two contributions to the energy separately. 

\subsubsection{Angular Effective Hamiltonian}

The Berry curvature of the lower energy band is peaked around ${\bf p}={\bf 0}$ where the band-gap is smallest (Fig. \ref{fig:scales}). Increasing the Zeeman field widens the band gap and spreads out the Berry curvature in momentum space. The ring of minima defines a closed contour in momentum space. If adiabatically transported around the ring, a single particle gains a geometrical Berry phase\cite{berry}:
\begin{eqnarray}
\Phi &=&  \int_S \Omega_-({\bf p}) dS = - \pi   \left(  \mbox{sign}(\Delta) - \frac{1}{\zeta}\right) ,  \label{eq:flux}
\end{eqnarray}
where the surface $S$ is bounded by the ring. This gauge-invariant phase can be recognised as the analogue of the magnetic flux, $e\Phi =  e\int_S B({\bf r}) dS$, in momentum space. 

\begin{figure} [htdp]
\centering
 \resizebox{0.4\textwidth}{!}{\includegraphics*{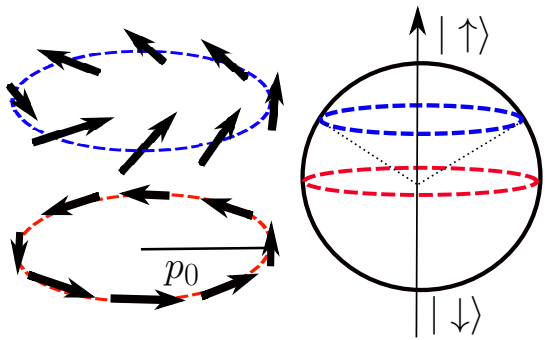}}  
\caption{(Color online) {\it LHS}: The orientation of the local spin vector for ({\it above}) intermediate and ({\it below}) large $\zeta$ around the ring of minima for $\Delta > 0$. {\it RHS} The mapping of the local spin vector onto the spin-1/2 Bloch sphere. As $\zeta \rightarrow \infty$, the spins at $p_0$ lie in the $xy$ plane, and the mapped path traverses the equator on the Bloch sphere. This corresponds to the maximum possible Berry phase, $|\Phi |=\pi$, for this model.} \label{fig:bloch}
\end{figure}

In the limit that $\Delta\rightarrow0$, the momentum-space magnetic flux, $\Phi$ is equal to $\pi$ and is concentrated at ${\bf p =0}$, where the two bands are degenerate\cite{cong-jun_unconventional_2011, sinha_trapped_2011, ghosh_trapped_2011, li_two-_2012, ramachandhran_half-quantum_2012, zhou_unconventional_2013}. When $\Delta \neq 0$, the flux can be tuned by varying the strength of the spin-orbit coupling or the Zeeman field. However, for a spin-$1/2$ particle, the flux is limited to $-\pi \leq \Phi < \pi$. To understand this, we recognise that the geometrical Berry phase arises from the rotation of the particle's spin as it travels around the ring of minima. Mapping the path around the ring onto the Bloch sphere (Fig. \ref{fig:bloch}), the Berry phase (\ref{eq:flux}) is equivalent to half of the solid angle enclosed by the path. For this model, the maximum possible value is $\pm \pi$, corresponding to the spins lying in the $xy$ plane when $\zeta \rightarrow \infty$. In Section \ref{sec:higher}, we shall discuss how this limitation may be overcome by using higher spin systems. 

The Berry phase can also be directly related to the Berry connection as: 
\begin{eqnarray}
\Phi &=& \oint \boldsymbol{\mathcal{A}} ({\bf p}) \cdot d {\bf l}  , \label{eq:cont}
\end{eqnarray}
where the line integral is around the ring of minima. The Berry connection of the particle confined to the ring is: 
\begin{eqnarray}
\boldsymbol{\mathcal{A}} (p_0, \varphi) &=& \frac{\Phi}{2 \pi p_0} \hat{\varphi}, \label{eq:berring}
\end{eqnarray}
for $\Delta > 0$. Alternatively this functional form could be derived directly from the spinor (\ref{eq:spinor}), by evaluating the Berry connection at ${\bf p}= (p_0, \varphi)$ and performing algebraic manipulations. 

We note that if the sign of the Zeeman field is reversed, $\Delta < 0$, the orientation of the local spin vector on the Bloch sphere begins from $|\downarrow \rangle$ as $\zeta$ increases (instead of from $|\uparrow \rangle$ as shown in Fig. \ref{fig:bloch}). However, the spinor  (\ref{eq:spinor}) has a singularity when $p=0$ and $\Delta < 0$, corresponding to the $|\downarrow \rangle$ state; this singularity must be removed for Eq. \ref{eq:cont} to hold. This can be done by gauge-transforming the spinor, multiplying it by a factor of $e^{-i \varphi}$, so that the singularity is instead at $|\uparrow \rangle$. Then the above form of the Berry connection (\ref{eq:berring}) can again be derived. 

From this Berry connection, we write the angular effective Hamiltonian (\ref{eq:effective2}) as: 
\begin{eqnarray}
\tilde{\mathcal{H}}_\varphi &=& \frac{\kappa}{2}
	\left(\frac{i}{p_0} \frac{\partial}{\partial \varphi} 
	+
	\frac{\Phi}{2 \pi p_0}
	\right)^2  , \label{eq:ring}
\end{eqnarray}
where all contributions from the lower band dispersion, $E_- ({\bf p})$ are included in the radial Hamiltonian below. As we shall now discuss, this is the momentum-space analogue of the Hamiltonian for a particle on a 1D real-space ring pierced by a magnetic flux. 

\subsubsection{A 1D Ring Pierced by Magnetic Flux}

In real space, the eigenspectrum and eigenstates of a single charged particle on a 1D ring pierced by a tuneable magnetic flux are well-known, and the flux has important physical consequences in both persistent currents and Aharonov-Bohm oscillations \cite{aharonov_1959, webb_1985}. The Hamiltonian of a particle on a 1D ring threaded by magnetic flux, $\Phi'$, is \cite{viefers}: 
\begin{eqnarray}
{\mathcal{H}}_\theta &=& 
	  \frac{1}{2M}
	\left(- \frac{i}{r_0} \frac{\partial}{\partial \theta} 
	- 
	\frac{e \Phi'}{2 \pi r_0}
	\right)^2,  \label{eq:ringr}
\end{eqnarray}
where $r_0$ is the radius of the ring and $(r, \theta)$ are polar coordinates in real space. Comparing Eqs. \ref{eq:ring} \& \ref{eq:ringr}, we see that these are analogous Hamiltonians with the roles of position and momentum reversed.  

The eigenstates of Eq. \ref{eq:ringr} in real space are \cite{viefers}:
\begin{eqnarray}
\psi_m = \frac{1}{\sqrt{2 \pi}} e^{ i m \theta} ,
\end{eqnarray}
while the energy spectrum is: 
\begin{eqnarray}
E_m' = \frac{1}{ 2M r_0^2 } \left( m - \frac{\Phi'}{\Phi'_0} \right)^2 , \label{eq:specring}
\end{eqnarray}
where we have introduced the magnetic flux quantum, $\Phi'_0 = 2 \pi / e$ (as we have set $\hbar=1$). The energy spectrum is parabolic in $m$, and periodic as the magnetic flux varies by $\Phi'_0$. 

We translate these results from the real-space Hamiltonian into momentum space. Then we find that the eigenstates of Eq. \ref{eq:ring} are: 
\begin{eqnarray}
G_m (\varphi) = \frac{1}{\sqrt{2 \pi}} e^{ i m \varphi} ,
\end{eqnarray}
while the energy spectrum is: 
\begin{eqnarray}
E_m = \frac{\kappa}{2 p_0^2} \left( m -\frac{\Phi}{\Phi_0} \right)^2 , 
\end{eqnarray}
and the analogue of the ``magnetic flux quantum": $\Phi_0 = 2 \pi $. In terms of our dimensionless parameters, this is: 
\begin{eqnarray}
\frac{E_m}{\omega} &=&  \frac{1}{2  \left(\frac{\zeta}{\chi} - \frac{1}{\zeta \chi} \right)} \left ( m +  \frac{1}{2}   \left(  \mbox{sign}(\Delta) - \frac{1}{\zeta}\right) \right)^2 . \label{eq:ring2}
\end{eqnarray}
The radius of the momentum-space ring varies as the flux is tuned. Consequently, the energy is not a unique function of $\Phi$, which it is in real space (\ref{eq:specring}), but depends on which parameter is used to tune the flux. For example, the flux may be tuned to $|\Phi/ \Phi_0 |\rightarrow 1/2$ by increasing the spin-orbit coupling strength: $\lambda^2 M \rightarrow \infty$. Then the radius of the momentum-space ring goes to infinity, and $E_m / \omega \rightarrow 0$. Alternatively, the flux could be increased by reducing the Zeeman field to zero. Then the momentum-space radius tends to a constant: $p_0 \rightarrow \sqrt{\zeta /\chi}$. The energy spectrum is then\cite{cong-jun_unconventional_2011, sinha_trapped_2011, ghosh_trapped_2011, li_two-_2012, ramachandhran_half-quantum_2012}: 
\begin{eqnarray}
 \frac{E_m}{\omega} &\xrightarrow[\Delta \rightarrow 0]& \frac{ \omega}{2 \lambda^2 M } \left ( m +  \frac{1}{2} \right)^2 , \label{eq:degen}
 \end{eqnarray}
where we have chosen $\mbox{sign}(\Delta)=1$ to avoid ambiguity when $\Delta=0$. This corresponds to choosing the spinor in the gauge of Eq. \ref{eq:spinor}. 

The single-particle ground states without a Zeeman field are the degenerate eigenstates with $m=0$ and $m=- 1$\cite{cong-jun_unconventional_2011}. Including a Zeeman field, breaks time-reversal symmetry and lifts this degeneracy; for $- 1/2 < \Phi/\Phi_0 < 1/2$, the ground state is always $m=0$\footnote{Without a Zeeman field, time-reversal symmetry can be broken instead by either spin-dependent interactions\cite{ramachandhran_half-quantum_2012} or beyond-mean-field fluctuations\cite{cong-jun_unconventional_2011, hu_spin-orbit_2012}}. As it is experimentally more relevant to consider $\Delta=0$ rather than $\lambda^2 M \rightarrow \infty$, we hereafter focus on tuning the flux via the Zeeman field.  

\subsubsection{Radial Effective Hamiltonian}

We assume that the particle is well-described by the properties of the ring of minima. Radially, we apply the effective mass approximation, expanding the energy bandstructure around this radius as: 
\begin{eqnarray}
E_{-} ({\bf p }) &\simeq& E_- (p_0  )  +  \frac{(p-p_0)^2 }{2 M^*} \nonumber\\
M^* & =& \frac{1}{(\partial^2 E_- ({\bf p}) / \partial ^2 { \bf p} )} \bigg|_{|{\bf p}|=p_0} = \frac{M} {1 - 1/ \zeta^2}.  \label{eq:massive}
\end{eqnarray}
Under these assumptions, $f(p)$ obeys a radial effective Hamiltonian (\ref{eq:effective2}): 
\begin{eqnarray}
\tilde{\mathcal{H}}_p \simeq - \frac{\kappa }{2} \frac{\partial^2}{\partial p^2} + \frac{1}{2 M^*} ( p-p_0)^2 +  E_- ( p_0  ) , \label{eq:radial}
\end{eqnarray}
as the radial component of Berry connection is zero. This is the Hamiltonian of a 1D simple harmonic oscillator in momentum space\cite{zhou_unconventional_2013, chen_2014}. The eigenstates are therefore: 
\begin{eqnarray}
f_n (p) = \frac{1}{\sqrt{2^n n!}} \left(\frac{1}{\pi p_1^2}\right)^{1/4}  e^{ - (p - p_0)^2  /2 p_{1}^2  } H_n (p / p_1)
\end{eqnarray}
where $H_n$ are the Hermite polynomials and the characteristic momentum scale is: 
\begin{eqnarray}
p_1 = \frac{1}{l_0}{ \left( \frac{M^*}{M}\right)}^{1/4}. \label{eq:width}
\end{eqnarray}
As $\zeta$ increases, the effective mass, $M^*$, tends towards the bare particle mass, $M$, from above (\ref{eq:massive}), and the characteristic localisation of the wave function in momentum space: $p_1 \rightarrow 1/ l_0 $. This dependence on the harmonic oscillator length is because a weaker harmonic trap in real space corresponds to a smaller kinetic energy in momentum space, and hence a more localised wave function in momentum space. 

The spectrum of (\ref{eq:radial}) is the well-known ladder of harmonic oscillator states: 
\begin{eqnarray}
E_n &=& \left(n + \frac{1}{2} \right) \sqrt{\frac{\kappa}{M^*}} +  E_- ( p_0  ) ,
\end{eqnarray}
which, in terms of our dimensionless parameters, is:
\begin{eqnarray}
\frac{E_n}{\omega}&=& \left(n + \frac{1}{2} \right) \sqrt{1 -  \frac{1 }{\zeta^2 }}  - \frac{1}{2} \left(\frac{\zeta}{ \chi} + \frac{1}{\zeta \chi}\right).  \label{eq:totalradial}
\end{eqnarray}
This reduces, in the limit of $\Delta \rightarrow 0$, to the previously known result\cite{cong-jun_unconventional_2011, sinha_trapped_2011, ghosh_trapped_2011, li_two-_2012, ramachandhran_half-quantum_2012, zhou_unconventional_2013, chen_2014}.

\subsection{Comparison of Analytical and \\Numerical Eigenspectra}

We now compare the analytical eigenspectrum in momentum space, with the numerical calculations outlined in Section \ref{sec:numerics}, based on the full Hamiltonian (\ref{eq:full}). 

The total energy, $E_{m,n} = E_m + E_n$, in the ring minima regime is found from adding Eqs. \ref{eq:ring2} \& \ref{eq:totalradial}: 
\begin{eqnarray}
\frac{E_{m,n}}{\omega} &=&- \frac{1}{2} \left(\frac{\zeta}{ \chi} + \frac{1}{\zeta \chi} \right) +  \left(n + \frac{1}{2} \right)  \sqrt{1 -  \frac{1 }{\zeta^2 }}   \nonumber \\
&& + \frac{1}{2  \left(\frac{\zeta}{\chi} - \frac{1}{\zeta \chi} \right)} \left ( m +  \frac{1}{2} \left(   \mbox{sign}(\Delta) - \frac{1}{\zeta}\right) \right)^2 \label{eq:tot}
\end{eqnarray}
For large $\zeta \gg 1$, the spacing in $n$ of the radial harmonic oscillator levels dominates over the energy splitting in $m$. This separation of scales is observed in the numerical energy spectrum in Figure \ref{fig:overview}. 

\begin{figure} [htdp]
\centering
$
\begin{array}{c}
 \resizebox{0.42\textwidth}{!}{\includegraphics*{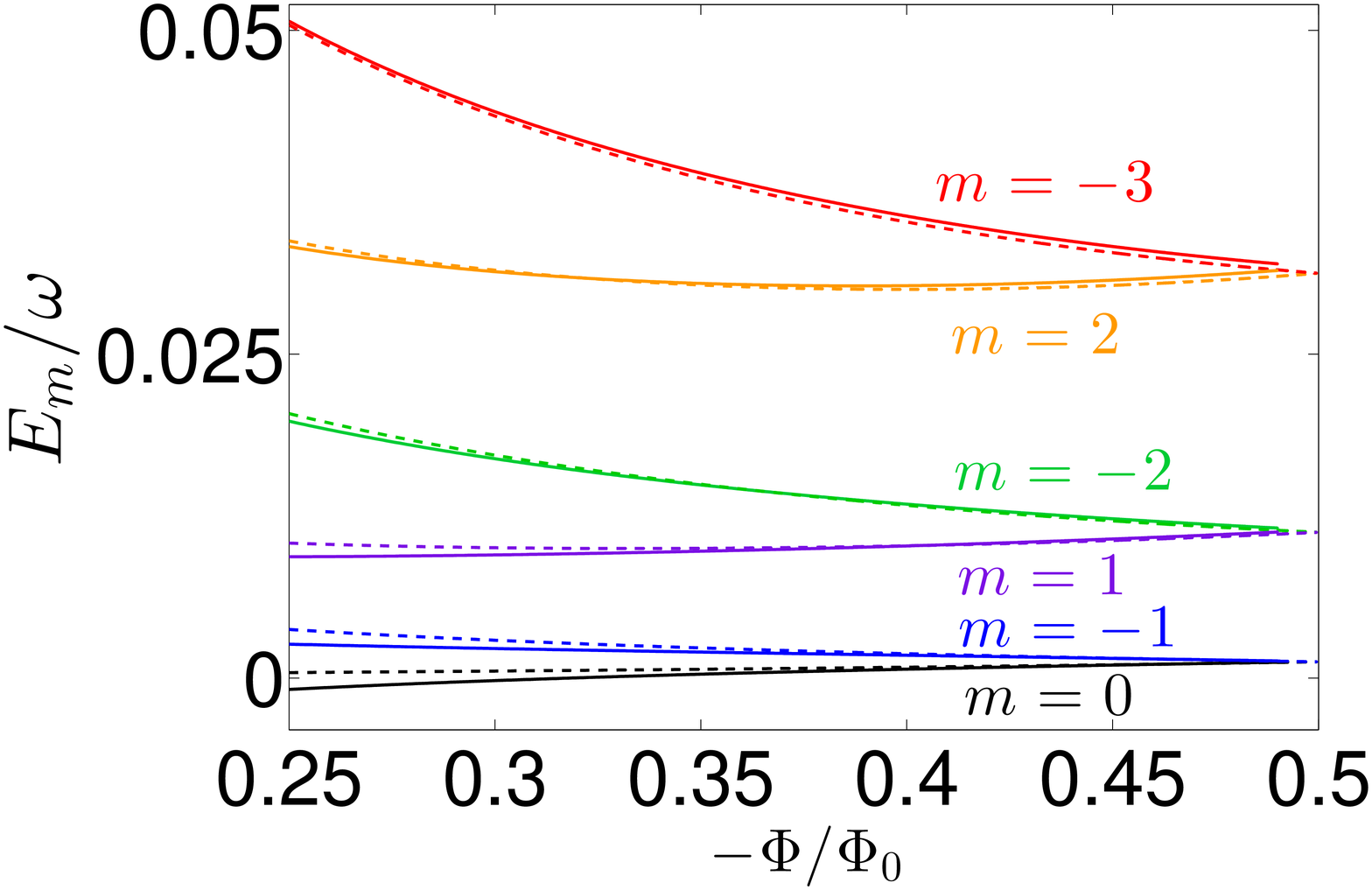}}  
  \end{array}$
\caption{(Color online) A segment of the energy spectrum as a function of the momentum-space magnetic flux, tuned via the Zeeman field. In the limit $\Delta \rightarrow 0$, $\Phi/ \Phi_0 \rightarrow - 1/2$, for $\zeta /\chi  = 100$. The dashed lines indicate the analytical predictions from Eq. \ref{eq:ring2}. The solid lines are extracted from numerics by calculating computationally the total energy and then subtracting the theoretical radial energy, $E_n / \omega$ (Eq. \ref{eq:totalradial}).} \label{fig:delta}
\end{figure}

Guided by the nearly flat energy dispersion in $m$, the single particle Hamiltonian without a Zeeman field has previously been mapped to a 2D Landau level Hamiltonian at large spin-orbit-coupling strength\cite{li_two-_2012, zhou_unconventional_2013}. The radial quantum number serves as the Landau level index, and the $n=0$ manifold has been termed the lowest Landau level. This description is reasonable provided that the angular momentum is small compared to $\zeta/\chi$. 

In Figure \ref{fig:delta}, we separate out the angular energy from the numerical results, in order to compare numerics with the energy spectrum of a momentum-space ring pierced by magnetic flux (\ref{eq:ring2}). We tune the flux by varying $\Delta$, while keeping $\zeta /\chi$ constant and large; the $\Delta=0$ limit is  captured by $-\Phi / \Phi_0 = 1/2$. As can be seen, there is excellent agreement between numerics and analytics at high flux for these parameters. As flux increases, $\zeta$ becomes larger, as does, consequently, the energy barrier in the centre of the ring, $ \Delta E = [E_-({ 0})- E_-({ p_0})]$. This barrier should be large compared with the harmonic trapping energy to justify our approximation that the particle is confined at the bottom of the ring. This approximation, and hence also the single-band approximation, improves as $\zeta$ increases and $\zeta/\chi$ increases. 
 
\subsection{Persistent Currents}

A remarkable feature of a quantum ring (or cylinder) threaded by a magnetic flux is the possibility of persistent charge currents, even in the ground state. These currents are an equilibrium property of the single-particle eigenstates; they flow without dissipation and reflect the phase coherence of the electron wave function around the ring. In bulk superconductors, macroscopic coherence is a key attribute of the superconducting wave function and persistent currents have been an important area of research since the 1960s\cite{deaver_1961, byers_1961, onsager_1961}. In recent years, this study has naturally been extended to Bose-Einstein condensates, where superfluid persistent mass currents in ring traps have also been experimentally studied by rotating or stirring the atomic cloud\cite{ryu_2007, ramanathan_2011, moulder_2012, beattie_2013}. 

Perhaps more surprisingly, persistent currents can exist in resistive rings, provided the phase coherence length is larger than both the elastic mean-free path and the circumference of the ring\cite{buttiker1983josephson, cheung1989persistent}. In these systems, persistent currents initially proved challenging to study experimentally because of decoherence from inelastic scattering and because of the necessity of using indirect experimental probes in order to preserve phase coherence of the electrons. However, persistent currents have now been extensively investigated in both mesoscopic metal rings\cite{levy_1990, chandrasekhar1991magnetic, jariwala_2001, deblock2002diamagnetic, bluhm_2009, Bleszynski_2009} and semiconductor rings\cite{mailly_1993,lorke_2000, kleemans_2007}. 

The phenomenology of persistent currents is further enriched by the inclusion of spin. In an inhomogeneous magnetic field, the spin of an electron with either spin-orbit coupling or a Zeeman interaction will rotate around the ring, and the electron can gain a spin Berry phase\cite{berry}.  This geometrical phase can be controlled by engineering the form of the magnetic field, with possible future applications in spintronic devices\cite{loss_1990, frustaglia, nagasawa2013control}. The spin Berry phase can be re-expressed in terms of a spin-dependent artificial magnetic gauge potential, which can generate persistent charge and spin currents around the real-space ring \cite{loss_1990, loss_persistent_1992, splettstoesser2003persistent, sun_2007, sun_2008}. (While spin-orbit coupling alone is sufficient to generate a real-space Berry phase around a ring, only a persistent spin current is produced as time-reversal symmetry is not broken \cite{sun_2007, sun_2008}.) Persistent spin currents of bosonic excitations have also been theoretically studied in Heisenberg rings\cite{schutz_2003}, and it has been proposed that persistent mass and spin currents may be created in ultracold gases using optically generated artificial magnetic fields\cite{song_persistent_2009, kanamoto_superpositions_2013}. 

We demonstrate that this physics is even more general than hitherto studied. Just as a real-space Berry phase or magnetic flux generates real-space currents, here we show how a momentum-space Berry phase can lead to eigenstates with the analogue of persistent currents {\it in  momentum space}. 

\subsubsection{Persistent Currents in Momentum Space}   \label{sec:currentmom}

In real space, the persistent current around a 1D ring pierced by magnetic flux can be calculated from the expectation value of the azimuthal velocity for a single particle: $I_{\theta} = \langle {\dot{r}}_\theta \rangle / 2 \pi r_0$. Due to the presence of the magnetic vector potential, the velocity operator must be defined with care from the magnetic Hamiltonian (\ref{eq:ringr}) as: 
\begin{eqnarray}
\dot{r}_\theta = - \frac{1}{e}\frac{\partial \mathcal{H}_\theta}{\partial {A}_\theta} =  \frac{1}{M r_0}  \left(-  i  \frac{ \partial }{\partial \theta} - \frac{e\Phi'}{\Phi'_0} \right) .
\end{eqnarray}
The average persistent current is then: 
\begin{eqnarray}
I_{\theta} =\frac{ \langle {\dot{r}}_\theta \rangle }{ 2 \pi r_0 }=  \frac{1}{2 \pi M r_0^2}  \left(m - \frac{e\Phi'}{\Phi'_0} \right) .  
\end{eqnarray}

{\it Analytical Calculation from the Effective Hamiltonian} --By analogy in momentum space, we consider the operator defined from the effective Hamiltonian (\ref{eq:ring}) as: 
\begin{eqnarray}
 \dot{p}_{\varphi} = - \frac{\partial \mathcal{ \tilde{H}_\varphi}}{\partial \mathcal{A}_\varphi}=  
- i \frac{\kappa}{p_0}   \frac{ \partial }{\partial \varphi} - \frac{\kappa}{p_0} \frac{\Phi}{\Phi_0} , \label{eq:operator}
\end{eqnarray}
where we have used that only the angular effective Hamiltonian depends on the angular Berry connection.

We calculate the expectation value of (\ref{eq:operator}) with respect to the low-energy eigenstates. We assume that $\zeta$ is large, so that the single particle is always in the $n=0$ radial ground state, then the eigenstate (\ref{eq:eigenstate}) can be written as: 
\begin{eqnarray}
\psi_{0m} ({\bf p}) &\simeq& G_m (\varphi ) \frac{f_0(p)}{\sqrt{p} } \nonumber \\
&=&  \left(\frac{1}{\sqrt{2 \pi^{3/2}  p_1 p}}\right) e^{ i m \varphi}     e^{ - (p - p_0)^2  /2 p_{1}^2  }  \label{eq:loweb}
\end{eqnarray} 
Calculating the expectation value of (\ref{eq:operator}) with respect to these states, we find: 
\begin{eqnarray}
\langle \dot{p}_{\varphi} \rangle = 
  \frac{\kappa}{p_0}   \left( m -  \frac{\Phi}{\Phi_0} \right) ,    \label{eq:100}
\end{eqnarray}
with a ``current" around the momentum-space ring: 
\begin{eqnarray}
I_{\varphi}& =& \frac{\langle \dot{p}_{\varphi}  \rangle}{2 \pi p_0} = 
 \frac{\kappa}{2 \pi p_0^2}  \left( m - \frac{\Phi}{\Phi_0} \right) .\label{eq:vel}
 \end{eqnarray}
As in real space, the ground state, $m=0$, supports a nonzero equilibrium ``current" that is directly proportional to the momentum-space magnetic flux threading the ring. When $\Delta =0$, the system is time-reversal symmetric and the ``current" in the $m=0$ state is equal and opposite to that in the $m=-1$ degenerate state. 

The role of the Berry connection in the effective Hamiltonian is to capture the behaviour of the spinor wave function. From $\eta_-({\bf p})$, we can understand the physical basis of the persistent current in momentum space. In the ring minima regime, $N_1=(1/2) (1 + 1/\zeta)$ of the particles are in one spin state carrying zero units of azimuthal angular momentum when $m=0$, while $N_2=(1/2)(1 - 1/\zeta)$ of the particles are in the other spin state, carrying one unit (\ref{eq:spinor}). The addition of these two contributions underlies Eq. \ref{eq:100} derived above.  

\begin{figure} [!]
\centering
 $
\begin{array}{c}
 \resizebox{0.4\textwidth}{!}{\includegraphics*{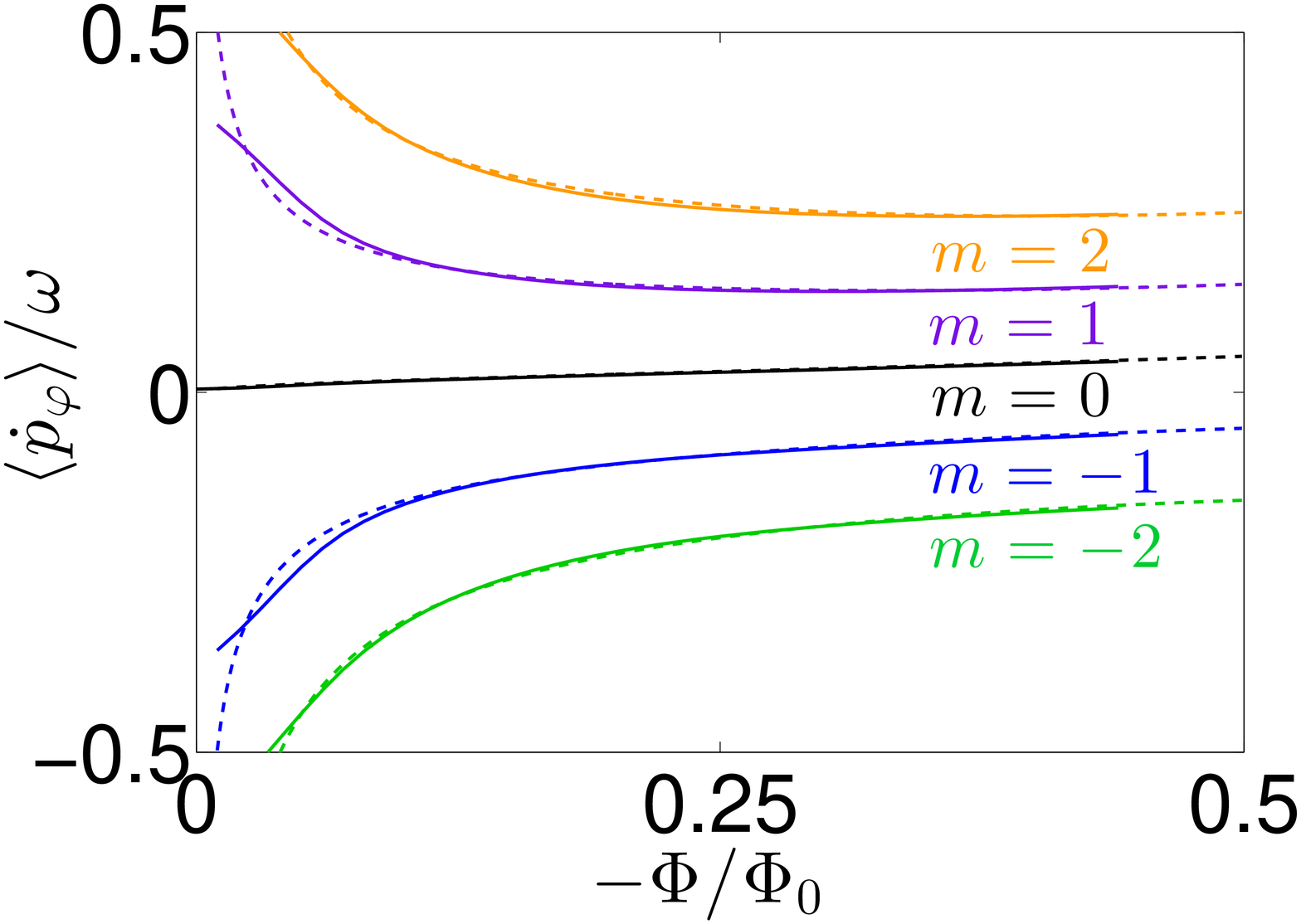}}  \\
\end{array} $
\caption{(Color online) The azimuthal ``velocity" in momentum space as a function of the flux for the five lowest energy states, $-2 \leq m \leq 2$ and $n=0$. The flux is varied by tuning the Zeeman field, $\Delta$, while holding the ratio $\zeta/\chi =100$ fixed. Dashed lines are analytical, while solid lines are numerical results.} \label{fig:andflux}
\end{figure}
{\it Numerical Calculation from the Full Hamiltonian} -- To compare this with numerical calculations, we must consider the relevant operators defined from the full Hamiltonian (\ref{eq:full}). To do so, we consider adding a fictitious momentum-space gauge potential, $\mathcal{A}_\varphi({\bf p})$, in the full Hamiltonian in momentum space and then taking: 
\begin{eqnarray}
 \dot{p}^F_{\varphi} &=& - \frac{\partial \mathcal{ {H}}}{\partial \mathcal{A}_\varphi} \bigg|_{\mathcal{A}_\varphi=0} \nonumber \\ &=& 
- \frac{\partial}{\partial A_\varphi} \left( \mathcal{H}_0 - \frac{1}{2} \kappa \frac{\partial^2}{\partial p^2} + \frac{1}{2} \kappa \left( \frac{i}{p} \frac{\partial}{\partial \varphi} + A_\varphi \right)^2 \hat{1} \right)\bigg|_{\mathcal{A}_\varphi=0} \nonumber \\
&=& - i \frac{\kappa}{p} \frac{\partial}{\partial \varphi} \hat{1} \label{eq:rate}
\end{eqnarray}
We calculate the expectation of this operator with respect to the momentum-space wave function of the full Hamiltonian. 

In Figure \ref{fig:andflux}, we compare the effective Hamiltonian approach (\ref{eq:100}) with numerics. As can be seen, the numerical and analytical results are in excellent agreement for large $\zeta /\chi$ above $|\Phi / \Phi_0 |\simeq 0.1$. In this regime, the harmonic trapping strength is small compared with the energy barrier in the centre of the trap and the approximations made analytically are valid. 

\subsubsection{Persistent Spin Currents in Momentum Space}   \label{sec:currentmom}

As mentioned above, when the spin of a particle rotates around a real-space ring, the spin Berry phase can lead to a persistent spin current in addition to the usual persistent current\cite{loss_1990, loss_persistent_1992, splettstoesser2003persistent, sun_2007, sun_2008}. In momentum space, by analogy,  the momentum-space Berry phase (Fig. \ref{fig:bloch}) can generate persistent spin ``currents" around the momentum-space ring.

To proceed, we must include the spin degree of freedom explicitly. We can approximate the low energy momentum-space eigenstate of the full Hamiltonian as: 
\begin{eqnarray}
\Psi_{0m} ({\bf p}) &\simeq&  \psi_{0m} ({\bf p}) \eta_- ({\bf p}) \label{eq:Psi}
\end{eqnarray}
for $\Delta >0$, where $ \psi_{0m} $ is the expansion co-efficient  (\ref{eq:loweb}) satisfying the effective Hamiltonian and $ \eta_- ({\bf p})$ is the spinor of the Hamiltonian in the absence of the harmonic trap $\mathcal{H}_0$. 

In general, a proper definition of the spin current density is challenging as the spin is not conserved in the presence of spin-orbit coupling. This issue has been extensively debated in the literature for real space (see for example Ref.~\onlinecite{sonin2010spin} and references therein), where different definitions have been suggested. We note that in this model, the momentum dependence of the local spin vector along $z$ is:
\begin{eqnarray}
\Psi_{0m}^\dagger ({\bf p}) \hat{\sigma}_z \Psi_{0m}({\bf p}) \propto \frac{1}{p \sqrt{p^2 \lambda^2 +\Delta^2}}  e^{ - (p-p_0) / 2p_1^2}
\end{eqnarray}
which is independent of the polar angle $\varphi$. This means that the $z$-component of the spin is constant at fixed radius $p$, such as around the ring of minima. We define a 1D azimuthal spin current density, $J^z_\varphi$, as the evaluation of the operator $\dot{p}^F_\varphi \hat{\sigma}_z $, with respect to the 1D eigenstates at the ring radius $G_m (\varphi) \eta_- (p_0, \varphi)$ for $\Delta >0$. We integrate around the ring of minima to find: 
\begin{eqnarray}
\int_0^{2 \pi} J^z_\varphi  d\varphi = \frac{\kappa }{ 2 p_0 }\left(m \left(1 + \frac{1}{\zeta}\right) - (m+1) \left(1 - \frac{1}{\zeta}\right)\right) \nonumber 
\end{eqnarray}
This has a clear physical interpretation in the same terms as the persistent current discussed above: $N_1=(1/2) (1 + 1/\zeta)$ of the particles are spin-up, with $m$ units of azimuthal angular momentum, while $N_2=(1/2)(1 - 1/\zeta)$ of the particles are spin-down, carrying $m+1$ units. 

We repeat the above calculation for $\Delta <0$, where the 1D eigenstates are $G_m (\varphi) e^{ - i \varphi }\eta_- (p_0, \varphi)$. (The gauge transformation is required to derive the chosen form of the Berry connection (\ref{eq:berring}) as discussed above.) The 1D spin current is now: 
\begin{eqnarray}
\int_0^{2 \pi} J^{z}_\varphi  d\varphi = \frac{\kappa }{ 2 p_0 }\left((m-1) \left(1 - \frac{1}{\zeta}\right) - m \left(1 + \frac{1}{\zeta}\right)\right). \nonumber 
\end{eqnarray}
This is because reversing the sign of $\Delta$ also flips the sign of $\zeta = \lambda^2 M / \Delta$. Now, $N_2=(1/2) (1 - 1/\zeta)$ of the particles are spin-up, carrying $m-1$ units of azimuthal angular momentum around the ring, while  $N_1=(1/2)(1+ 1/\zeta)$ of the particles are spin-down, carrying $m$ units.

\subsubsection{Consequences of Momentum-Space Persistent Currents in Real Space}

Persistent currents around the momentum-space ring will also have physical consequences in real space. the analogy with charged particles in mesoscopic real-space rings, this would correspond to the momentum-space behaviour of the system which has been little studied. 

{\it Real-Space Magnetic Moment--} We begin from the real-space azimuthal velocity operator, $\hat{v}_\theta$, where $\theta$ is the polar angle in real space. This must be defined with care as the spin-orbit coupling acts as an additional gauge field, modifying the physical velocity. We introduce a fictitious gauge potential, $A_\theta$, into the full Hamiltonian (\ref{eq:full}), in real space, and derive the velocity operator as: 
\begin{eqnarray}
 \hat{v}_\theta &=&- \frac{1}{e}\frac{\partial H}{\partial A_\theta} \bigg|_{A_\theta=0}= - \frac{i}{M}\frac{1}{ r} \frac{\partial}{\partial \theta} \hat{1} - \lambda \hat{\sigma}_r , \nonumber \\
 \hat{\sigma}_r &=&\left(\begin{array}{cc}
  0 & e^{ - i \theta} \\e^{  i \theta}  & 0 
\end{array} \right) ,
\end{eqnarray}
where we have introduce the radial Pauli matrix $ \hat{\sigma}_r$. The first term can be called the ``kinetic" contribution, while the second term will be referred to as the ``spin-orbit" contribution. We calculate the real-space magnetic moment as:
\begin{eqnarray}
 \mu_z&\propto& M  \mu_B \langle r \hat{v}_\theta  \rangle  \label{eq:mu}
\end{eqnarray}
where $\mu_B = 1/ (2 M_e)$ is the Bohr magneton, $M_e$ is the electron mass and we have taken $\hbar=e=1$.  

To proceed, we Fourier-transform the real-space operators into momentum space: 
\begin{eqnarray}
r \hat{v}_{\theta} &&=- \frac{i }{M}\frac{\partial}{\partial \varphi} \hat{1}
\nonumber \\ &&
 - \lambda  \left(
\begin{array} {cc} 0 
  & e^{ - i \varphi} \left( \frac{1}{p} \frac{\partial}{\partial \varphi} + i \frac{\partial}{\partial p}\right) \\
 e^{ i \varphi} \left( - \frac{1}{p} \frac{\partial}{\partial \varphi} + i \frac{\partial}{\partial p}\right) & 0 \end{array} \right)  \label{eq:moop}
 \end{eqnarray}
where $\varphi$ is the polar angle in momentum space, and the first (second) term is the kinetic (spin-orbit) term. 

To calculate the kinetic contribution, we take the expectation value of the operator with respect to the wave function (\ref{eq:Psi}): 
\begin{eqnarray}
-i  \langle \Psi_{0m}  | \frac{\partial}{\partial \varphi} \hat{1} |  \Psi_{0m} \rangle  =  \int_0^\infty dp && \frac{-\Delta + (2m+1)\sqrt{p^2 \lambda^2 + \Delta^2}}{2 p_1  \sqrt{\pi}  \sqrt{p^2 \lambda^2 + \Delta^2} }  \nonumber \\
 && \times  e^{ - (p - p_0)^2  / p_{1}^2  }   .
 \end{eqnarray}
In the limit that the wave function is strongly localised radially around $p_0$, i.e. when $p_1 \rightarrow 0$ as the flux is large or the harmonic trap is weak (Eq. \ref{eq:width}), we can approximate this integral using the Laplace method as:
 \begin{eqnarray}
-i  \langle \Psi_{0m}  |  \frac{\partial}{\partial \varphi} \hat{1} |  \Psi_{0m} \rangle&&\simeq    
 - \frac{\Delta}{2 \sqrt{p_0^2 \lambda^2 + \Delta^2}} + \frac{(2m+1)}{2  }  \nonumber \\
 &&= \left(  m + \frac{1}{2} \left(1 - \frac{1}{\zeta} \right)\right) \label{eq:lmu}. 
 \end{eqnarray}
Repeating this calculation for the opposite Zeeman field, $\Delta <0$, we find: 
\begin{eqnarray}
-i \langle \Psi_{0m}'  | \frac{\partial}{\partial \varphi} \hat{1} |  \Psi'_{0m} \rangle \simeq  \left(  m - \frac{1}{2} \left(1 - \frac{1}{\zeta} \right)\right) . \label{eq:ang}
 \end{eqnarray}
 Combining these two results, and comparing with Eq. \ref{eq:flux}, we write this as:
\begin{eqnarray}
-i  \langle \frac{\partial}{\partial \varphi} \hat{1}  \rangle\simeq  \left(  m - \frac{\Phi}{\Phi_0}\right) \label{eq:lmu}. 
 \end{eqnarray}
This is similar to Eq. \ref{eq:100} calculated above as the Berry connection in the effective Hamiltonian is capturing the behaviour of the spinor wave function (which leads to the term $\propto \frac{\Phi}{\Phi_0}$). 

Similarly, we can calculate the spin-orbit contribution, again applying the Laplace approximation to the resulting integrals. We then find that: 
\begin{eqnarray}
- M \lambda \langle && r \hat{\sigma}_r   \rangle \simeq 
-   m - \mbox{sign}(\Delta) - \frac{\Phi}{\Phi_0} . 
 \end{eqnarray}
The total real-space magnetic moment is then:
\begin{eqnarray}
\mu_z \propto -  \frac{{\mu_B}}{{\zeta}} =   {\mu_B} \left(  - \mbox{sign}(\Delta)- 2 \frac{\Phi}{\Phi_0} \right), \label{eq:moment}
\end{eqnarray}
for all low-energy states, independent of the azimuthal quantum number, $m$. The magnetic moment is maximal in the limit of vanishing flux and tends to zero as $|{\Phi}/{\Phi_0}| \rightarrow 1/2$, corresponding to an increasing spin-orbit coupling and/or a vanishing Zeeman field. We find that this analytical result is in good agreement with the numerical calculations, for $|\Phi / \Phi_0| \geq 0.1$, for small values of $m$ and large values of $\zeta / \chi$.
 
The real-space magnetic moment (\ref{eq:moment}) can alternatively be derived semiclassically as the magnetic moment of a  wavepacket\cite{chang} confined to the 1D ring in momentum space. We note that the semiclassical magnetic moment was previously studied for a 2D electron gas in a Zeeman field with Rashba spin-orbit coupling in Ref.~\onlinecite{mishchenko}, where equilibrium real-space edge currents in the presence of a confining potential were also discussed.
 
{\it Real-Space Spin Density Profiles--} As introduced above, analogue persistent currents in momentum space can be understood in terms of the atom number in each spin component. In real space, this can be measured directly in the density profile of the two spin components. 

 \begin{figure} [!]
\centering
 $
\begin{array}{c}
 (a)\resizebox{0.45\textwidth}{!}{\includegraphics*{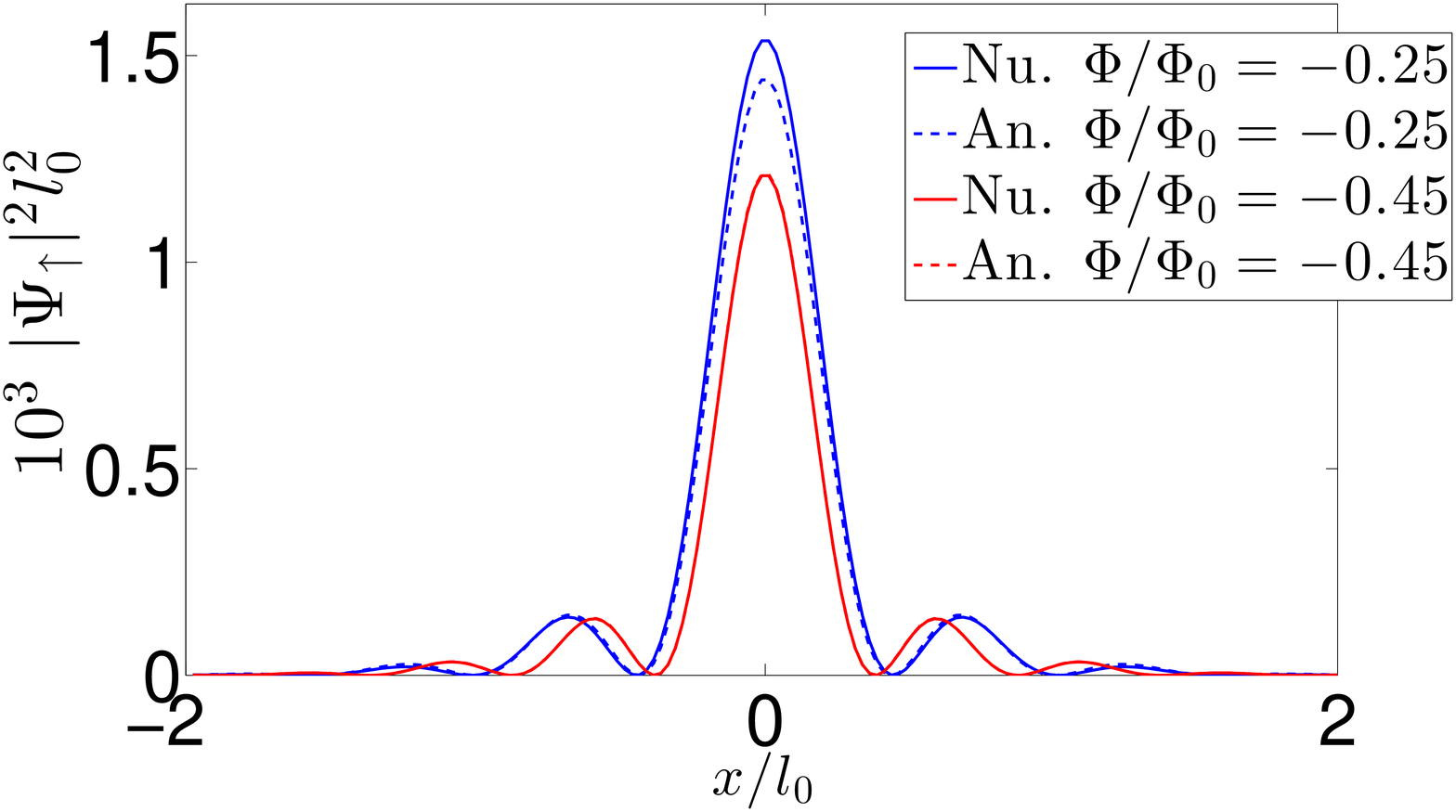}}  \\
 (b) \resizebox{0.45\textwidth}{!}{\includegraphics*{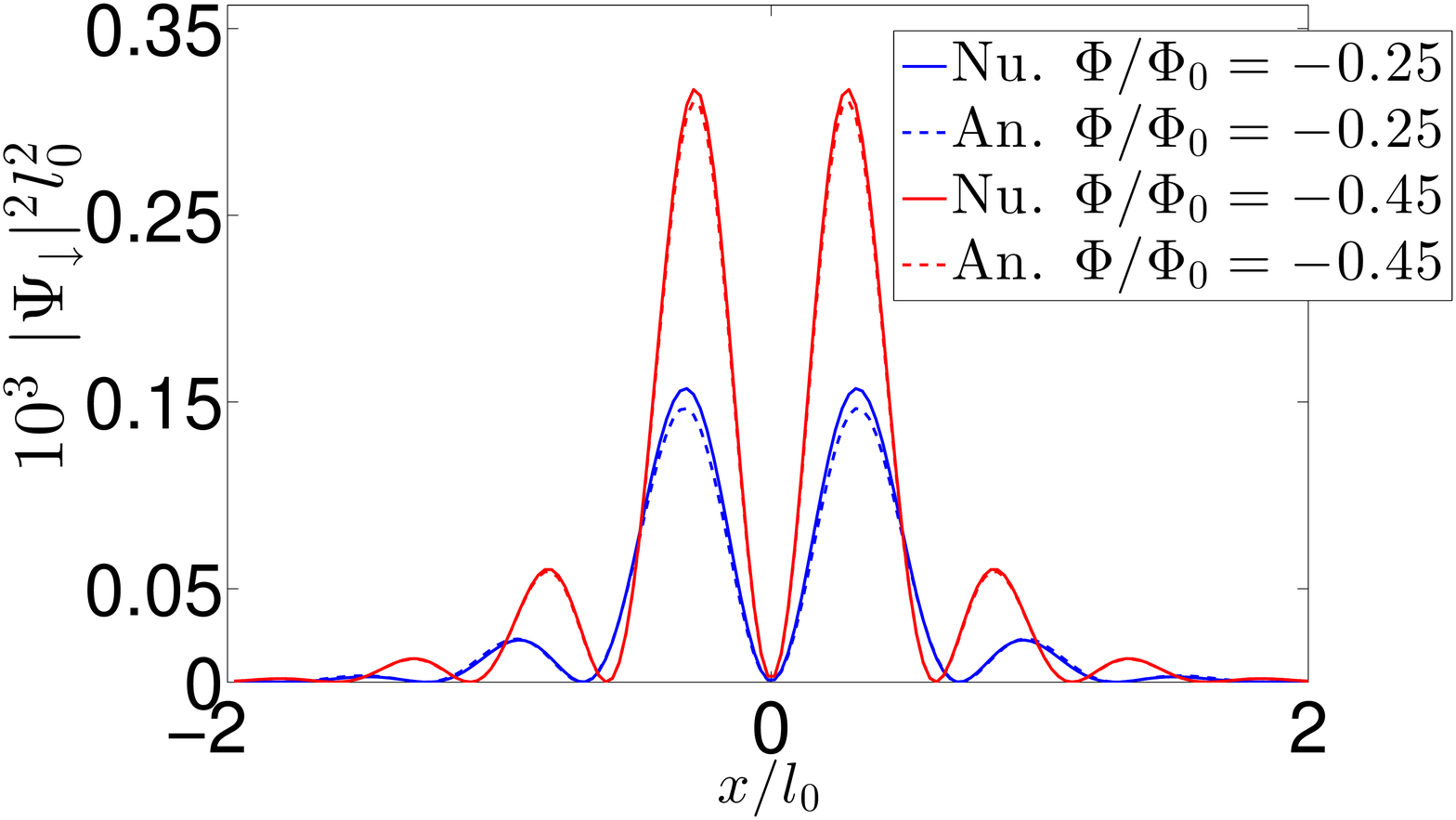}}
\end{array} $
\caption{(Color online) Comparison between Eq. \ref{eq:ansatz} (An.) and numerical (Nu.) real-space wave function for $\Phi / \Phi_0=-0.45$ and $\Phi/\Phi_0=-0.25$, with $\zeta/\chi=40$ and $m=0$. (a)The density of spin-up atoms and (b) the density of spin-down atoms for a cut along $y=0$. As the density profiles are rotationally symmetric, the full distribution can be found by rotation around the $z$-axis.} \label{fig:numericcom}
\end{figure}

To lowest order in the momentum, we can approximate the spinor as ${\bf \eta}_- ({\varphi, p})  \simeq \eta_- ({\varphi}, p_0)$; this approximation improves as $\Delta\rightarrow 0$. By neglecting the momentum dependence of the spinor, we can apply the approximate identity\cite{chen_2014}:
\begin{eqnarray}
\int d q \sqrt{q} e^{ - (q - p_0 l_0)^2 / 2 } J_m ( q r / l_0) \simeq \sqrt{2 \pi p_0 l_0}  e^{ -r^2 / 2 l_0^2} J_m ( p_0 r) \nonumber
\end{eqnarray}
where $J_m$ is a Bessel function of the first kind and where we have used that $p_1 \simeq 1/l_0$. This approximate identity is valid provided that $p_0 l_0 \gg m$, implying the ``thin-ring limit" where the harmonic trap strength is much larger than the single-particle angular energy, $E_m$\cite{chen_2014}. Under these conditions, Eq. \ref{eq:Psi} can be Fourier-transformed to find the approximate real-space wave function: 
\begin{eqnarray}
\Psi_{0, m}  ({\bf r}) &\simeq& i^m \frac{e^{ - r^2 / 2 l_{0}^2}} { \sqrt{2 l_0  \sqrt{\pi}}}    \sqrt{p_0 - \frac{p_0\Delta}{\sqrt{p_0^2 \lambda^2 +\Delta^2} }} 
\nonumber \\ &&
 \times   \left( \begin{array}{c}
 \frac{\Delta + \sqrt{p_0^2 \lambda^2 +\Delta^2}}{p_0 \lambda }e^{i m \theta}  J_{m} ( p_0  r) \\
  e^{i (m+1) \theta} J_{m+1} (p_0 r) 
\end{array} \right) .  \label{eq:ansatz}
\end{eqnarray}
This analytical result is compared with numerics from the full Hamiltonian in Figure \ref{fig:numericcom} for two values of the flux for the $m=0$ state. There is excellent agreement at high flux where the above approximations for the analytical wave function are most appropriate, but there is also good qualitative agreement with key features at lower flux values. 

The azimuthal angular momentum of the spin state is reflected in the real-space density profile through the Bessel function, $J_m$. In particular, if the spin component carries no angular momentum, there is a maximum in the density at ${\bf r}=0$, otherwise there is a node in the density at the real-space origin (Fig.  \ref{fig:numericcom}). 

The real-space wave function was previously studied for a vanishing Zeeman field, where the degenerate ground states, $m=0$ and $m=-1$, have been termed half-quantum vortex states\cite{cong-jun_unconventional_2011}. This description refers to half of the atoms being in an $s$-state with no units of angular momentum, while the other half are in a $p$-state with one unit. As the local spin vector winds radially outwards, the real-space wave function can also be described as a nontrivial topological Skyrmion-like spin texture\cite{cong-jun_unconventional_2011}. 

Introducing a small Zeeman field, tunes the number of particles between the two spin states (Fig. \ref{fig:numericcom}), without changing the qualitative features of the wave function. By imaging the real-space spin-up and spin-down density profiles, experiments could extract both the azimuthal angular momentum of each spin species as well as the proportion of particles in each component. In Sec.~\ref{sec:exp} we shall discuss further experimental ways to observe the effects we describe.

\section{Higher Spin Systems} \label{sec:higher}

In the ring minima regime, simple forms of the spin-orbit coupling, such as 2D Rashba or Dresselhaus for spin-1/2 particles, limit the tunability of the flux to: $-1/2 \leq \Phi/\Phi_0 < 1/2$. As mentioned above, this reflects the maximum value of the Berry phase possible in these models. To further extend the analogy with real-space magnetism, we show that the momentum-space flux  can be tuned over a larger range in higher spin systems. This may soon be experimentally relevant thanks to recent proposals for how spin-orbit coupling for higher spins might be generated using pulsed magnetic fields\cite{anderson_2013, xuueda} or the optical dressing of internal atomic states\cite{juzel_2010}. 

We discuss a generalised Rashba spin-orbit coupling for a particle in 2D with spin $F$ \cite{chen_2014}. The full Hamiltonian (\ref{eq:full}) becomes:
\begin{eqnarray}
\mathcal{H} &=& \mathcal{H}_0 + \frac{1}{2} \kappa {\bf r}^2 \hat{1},  \nonumber \\
\mathcal{H}_0 &=&  \frac{{\bf p}^2}{2M} \hat{1}+ \frac{\lambda}{F} (p_x \hat{F}_y - p_y \hat{F}_x) - \Delta \hat{F}_z  , \label{eq:higherh}
\end{eqnarray}
where $\hat{F}_{x,y,z}$ are the spin-F matrices along the $x$, $y$ and $z$ directions. In this Hamiltonian, we have neglected the quadratic Zeeman shift which may also be present for certain atomic species\cite{anderson_2013, xuueda}. 

The properties of this general model without a harmonic trap are given in Appendix A. This system has $2F+1$ bands, but the energy dispersion of the lowest band remains:
\begin{eqnarray}
E_{-} ({\bf p}) =  \frac{p^2}{ 2M } - \sqrt{p^2 \lambda^2 + \Delta^2}
\end{eqnarray}
with the ring of minima as before at $p_0 = \sqrt{ M^2 \lambda^2 - \Delta^2 /  \lambda^2}$ provided that $ M \lambda^2 / \Delta > 1$. The Berry curvature is\cite{berry}: 
\begin{eqnarray}
\Omega_{-} ({\bf p}) = -  F \frac{  \lambda^2 \Delta}{ ( \lambda^2 p^2 + \Delta^2)^{3/2}} , \label{eq:highberry}
\end{eqnarray}
and where Eq. \ref{eq:berry} is regained for $F=1/2$. 

\subsection{Single-Minimum Regime: Fock Darwin}

From Eq. \ref{eq:highberry}, the higher spin, $F$, multiplies the value of the Berry curvature at ${\bf p}={\bf 0}$. As a result, the ``cyclotron frequency" entering Eq. \ref{eq:enmcin} becomes:
\begin{eqnarray}
\frac{\omega_c }{\omega}= \frac{\kappa \Omega_0}{\omega}  =  - F \chi \zeta ,
\end{eqnarray} 
which scales the splitting between states with $(2 n + |m|)$. For sufficiently large spins, the effects of the Berry curvature will dominate over other terms in the energy spectrum. This will be directly observable in the dipole mode splitting of an ultracold atomic gas. 

In the limit of a very strong momentum-space magnetic field, the confinement from the energy dispersion becomes irrelevant and the free-particle behaviour dominates. The states may then be described as Landau bands\cite{chakraborty1999quantum} with an energy dispersion: 
\begin{eqnarray}
E&=&E_0  + \left(N + \frac{1}{2}\right) \omega_c
\end{eqnarray} 
where the Landau level number is $N = n + \frac{|m| - m }{2}$, and where we have used that $\omega_c \gg \omega_m$ in (\ref{eq:enfd}). 

The increase in $\Omega_0$ can also be seen in the momentum-space wave function, which has a characteristic length scale in momentum space (\ref{eq:length}):
\begin{eqnarray}
l_\Omega &=&  \sqrt{\frac{M \omega}{\hbar} }\left( \frac{M}{M^*} +  \frac{(M \omega \Omega_0)^2}{4}\right)^{-1/4} \nonumber \\
&=& \frac{1}{l_0} \left( ( 1 - \zeta) + F^2 \frac{(\zeta\chi)^2}{4}\right)^{-1/4}  .\label{eq:length2}
\end{eqnarray}
The wave function is more strongly localised as $F$ increases for given values of $\zeta$ and $\chi$, due to the increase in the strength of the momentum-space artificial magnetic field. One consequence of this is that the Fock-Darwin description is valid over a larger range of spin-orbit coupling strengths as the wave function is well-described by the properties of the band at ${\bf p}={\bf 0}$.  

\subsection{Ring-Minima Regime}

\begin{figure} [!]
\centering
 $
\begin{array}{c}
 (a)\resizebox{0.4\textwidth}{!}{\includegraphics*{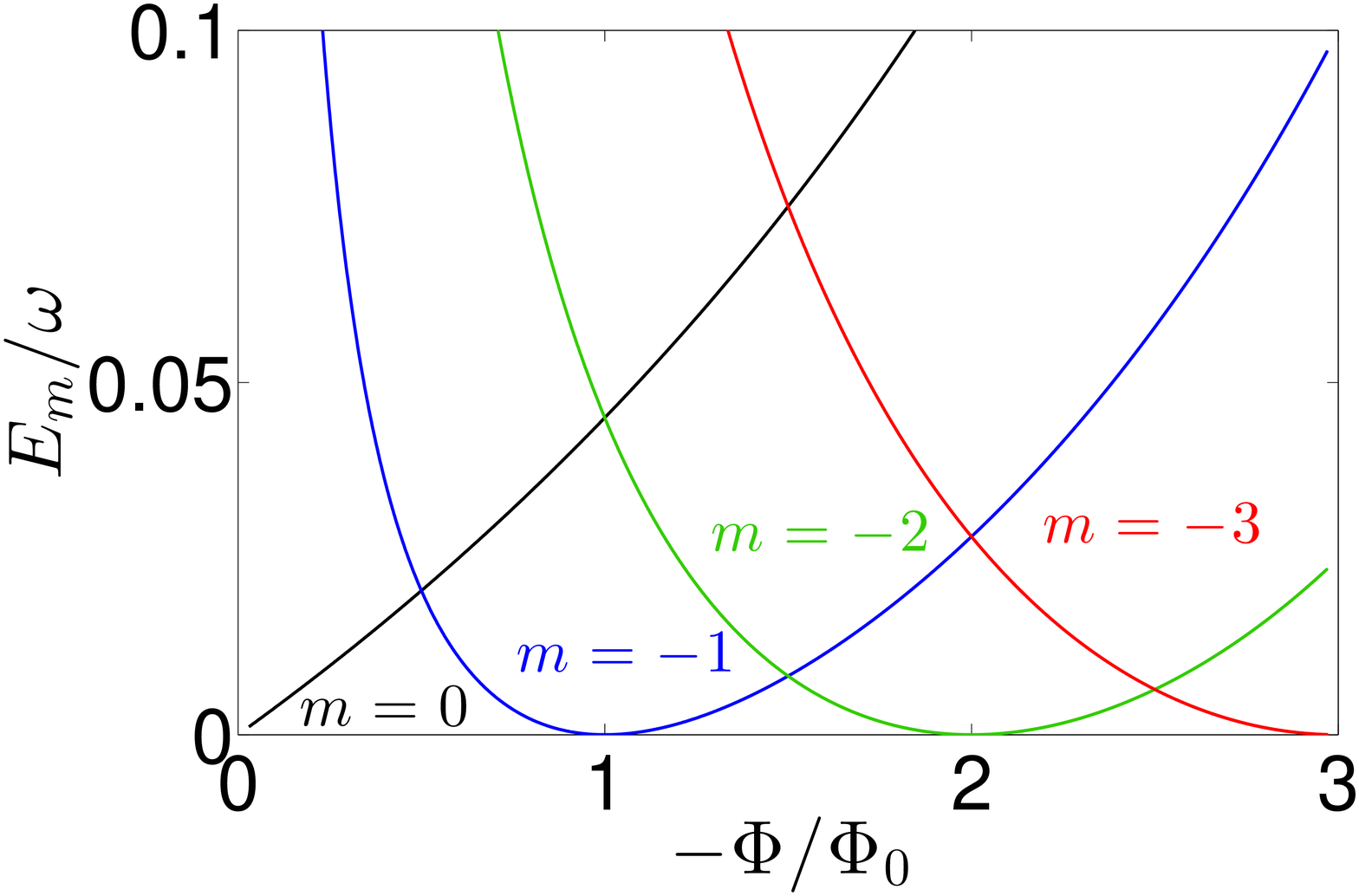}}  \\
 (b) \resizebox{0.41\textwidth}{!}{\includegraphics*{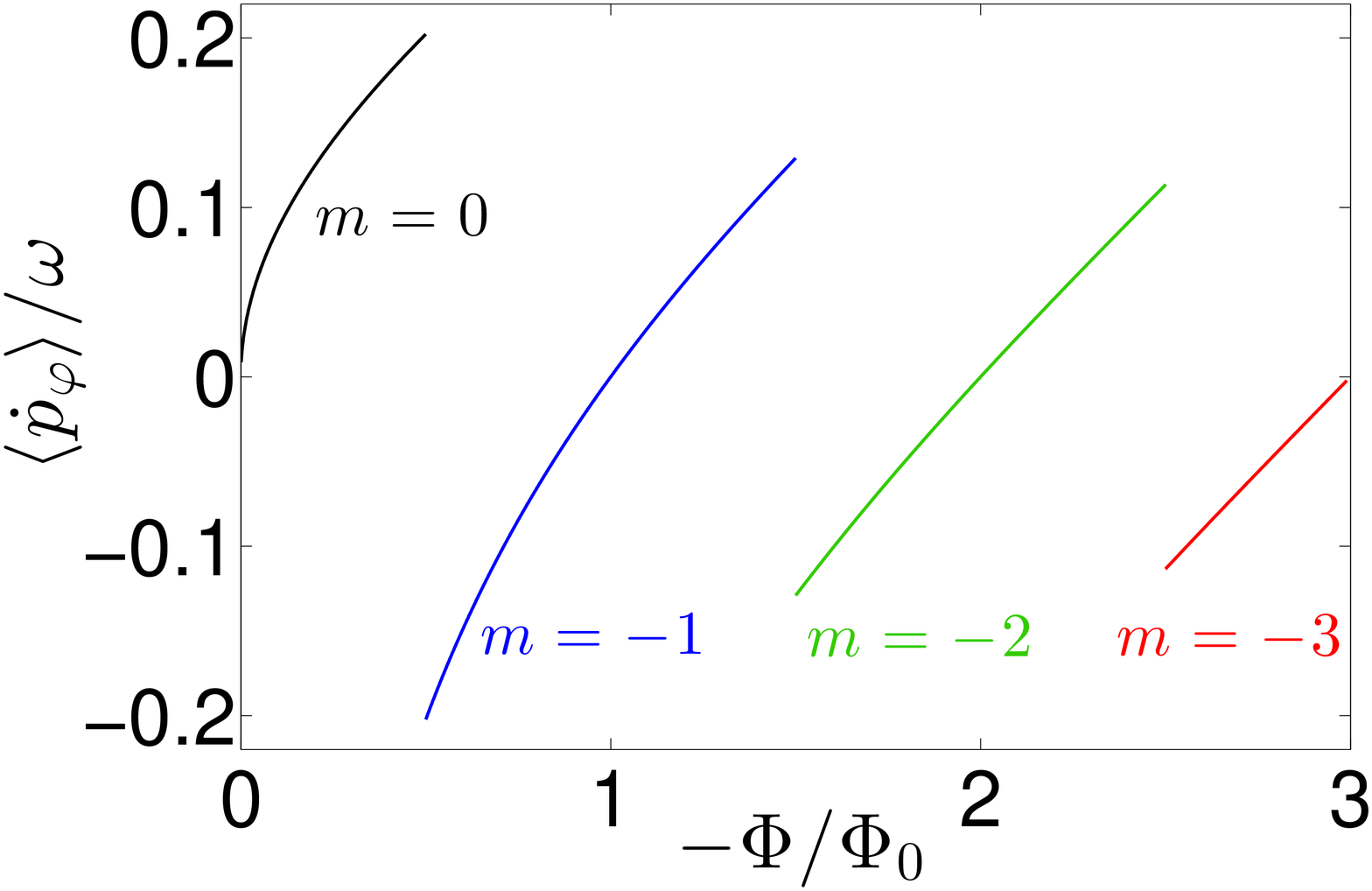}}  \\
\end{array} $
\caption{(Color online) (a) The analytical angular energy (\ref{eq:2sp}) for $\zeta / \chi =20$ and $F=3$, showing the successive ground states with different values of $m$. The degeneracies between states occur at half integer values of the flux. (b) The analytical ``velocity" for the ground state (\ref{eq:3sp}) for the same parameters.} \label{fig:higher}
\end{figure} 

In the ring minima regime, the momentum-space artificial magnetic flux  is: 
\begin{eqnarray}
\frac{\Phi}{\Phi_0} &=& \frac{2 \pi}{\Phi_0} \int_0^{p_0} \Omega_{-}(p)  p  dp = -F \left(   \mbox{sign}(\Delta)  - \frac{1}{\zeta}\right) \label{eq:fluxsp}
\end{eqnarray}
for the lowest band. This is because the Berry phase is $F$ times the solid angle enclosed by the evolution of the spin vector around a path (Fig. \ref{fig:bloch}) for a spin model of arbitrary spin\cite{berry}. In the higher-spin Hamiltonian (\ref{eq:higherh}), the path taken by the spin vector is independent of $F$, and the spin enters as a simple multiplicative factor into the flux. The range of the flux is now:  $-F \leq \Phi/\Phi_0 < F $. The higher spin also affects the angular energy (\ref{eq:ring2}): 
\begin{eqnarray}
\frac{E_m}{\omega} &=&  \frac{\kappa}{2 \omega p_0^2} \left( m -\frac{\Phi}{\Phi_0} \right)^2 \nonumber \\
&=& 
 \frac{1}{2  \left(\frac{\zeta}{\chi} - \frac{1}{\zeta \chi} \right)} \left ( m +  F  \left(  \mbox{sign}(\Delta)  - \frac{1}{\zeta}\right) \right)^2 ,  \label{eq:2sp}
\end{eqnarray}
%s
as demonstrated in Figure \ref{fig:higher}(a) for $F=3$. Now that the flux can be tuned past $|\Phi / \Phi_0|=1/2$, the successive transitions of the ground state between different values of $m$ are apparent. As for a mesoscopic normal ring, the transitions occur for half-integer flux while the minimum energies occur for integer flux. However, unlike a real-space ring, the energy is not perfectly periodic with $\Phi =2 \pi$ due to the variation in $p_0$ with the flux. 

In the special case that the Zeeman field vanishes, the energy is: 
\begin{eqnarray}
 \frac{E_m}{\omega} &\xrightarrow[\Delta \rightarrow 0]& \frac{\hbar \omega}{2 \lambda^2 M } \left ( m + F  \right)^2 ,
 \end{eqnarray}
where we have again chosen $\mbox{sign}(\Delta)=1$ in this limit. 

As previously studied for spin-1/2, a half-integer spin-orbit-coupled system without a Zeeman field has a non-trivial Berry phase of  $\pi$ modulo $2 \pi$, and all states are doubly degenerate. Conversely, for integer spin, the effect of $F$ is simply to relabel the integers $m$. This corresponds to a trivial Berry phase of $0$ modulo $2 \pi$. (The absence of a Berry phase for spin-1 systems has previously been noted in Ref.~\onlinecite{chen_2014}.) In this case, the ground state is always unique and has azimuthal angular momentum $m=- F$. 
 
Similarly, we can generalise the azimuthal ``velocity" to higher spins (\ref{eq:100}):
\begin{eqnarray}
\langle \dot{p}_{\varphi} \rangle = 
  \frac{\kappa}{p_0}   \left( m  +  F \left(   \mbox{sign}(\Delta)  - \frac{1}{\zeta}\right) \right) , \label{eq:3sp}
\end{eqnarray}
illustrated in Figure \ref{fig:higher}(b). As the ground state transitions between different states, the ``velocity" of the ground state jumps. This is analogous to the ``sawtooth" behaviour of the persistent current in the ground-state in mesoscopic real-space rings\cite{viefers}.

\subsection{ Single-Band Approximation}

The effective momentum-space magnetic Hamiltonian requires the validity of the single-band approximation. Here we discuss in detail this approximation in all regimes for a general spin-$F$ particle. (The conditions for a spin-1/2 particle that have been discussed in the above text are regained by setting $F=1/2$.) In the single band approximation, we assume that the harmonic trapping energy is much smaller than the band gap between the lowest and second lowest band. The second lowest band has energy: 
\begin{eqnarray}
E_{2} ({\bf p}) =  \frac{p^2}{ 2M } - \left( 1 - \frac{1}{F}\right) \sqrt{p^2 \lambda^2 + \Delta^2} . 
\end{eqnarray}
{\it Regime 1: }For $F= 1/2$ or $\zeta < \frac{F}{F-1}$, there is a single minimum in the second lowest band at ${\bf p}={\bf 0}$. The energy is:
\begin{eqnarray}
\frac{E_{2} (0)}{\omega} = - \left( 1 - \frac{1}{F}\right) \frac{1}{\chi} . 
\end{eqnarray}
Provided that $\zeta < 1$, the lowest band is also in the single minimum regime, with energy ${E_{-} (0)}/{\omega} = - 1 /\chi$. Comparing these energy scales, we see that the single band approximation is satisfied if $1 \gtrsim F \chi$. This requirement becomes harder to satisfy for higher spin systems and its experimental consequences will be discussed in more detail below.  

{\it Regime 2: }For either $F=1/2$ and $\zeta>1$ or for $F\geq1$ and $ 1 < \zeta < \frac{F}{F-1}$, the lower band will be in the ring minima regime while the second lowest band will have a single minimum at ${\bf p}={\bf 0}$. In the ring minima regime, we require that the harmonic trapping strength is small compared with the energy barrier in the centre of the ring. As this barrier is smaller than the band gap, $(E_{2} (0) - E_- (p_0)) \gtrsim (E_{-} (0) - E_- (p_0))$, the single-band approximation is automatically satisfied provided that the trap is small compared with the energy barrier: $(E_{-} (0) - E_- (p_0)) / \omega \lesssim 1$. This requires: 
\begin{eqnarray}
\chi \lesssim \frac{1}{2}\left( \zeta + \frac{1}{\zeta} \right) - 1  ,
\end{eqnarray}
which is independent of the spin, F. In the limit of a vanishing Zeeman field for a spin-1/2 particle, this condition is approximately $2 \lesssim \zeta / \chi$.  

{\it Regime 3: }For higher spin systems, when $ \frac{F}{F-1} < \zeta$, the second lowest band also has a ring of minima at: 
\begin{eqnarray}
p_2 = \sqrt{\frac{(F-1)^2}{F^2} M^2 \lambda^2  - \frac{\Delta^2}{\lambda^2}} , 
\end{eqnarray}
where the energy  is:
\begin{eqnarray}
\frac{E_{2} (p_2)}{\omega} = - \frac{1}{2\chi \zeta} - \frac{(F-1)^2}{2F^2} \frac{\zeta}{\chi} .
\end{eqnarray}
For large $\zeta$, this minimum energy can be lower than the energy of the lowest band at the origin. Then the band gap is smaller than the height of the barrier at the centre of the ring, $(E_{2} (p_2) - E_- (p_0)) \lesssim (E_{-} (0) - E_- (p_0))$, and the single-band approximation must be reinforced by ensuring that:  
\begin{eqnarray}
1 &\lesssim& \left(1 - \frac{(F-1)^2}{F^2} \right)  \frac{\zeta}{2 \chi} , \label{eq:regime3}
\end{eqnarray}
This requirement also becomes more stringent at larger $F$; for example, for $F=5$, this requires that $\zeta / \chi \gtrsim 5.5$. 

\section{Experimental Considerations} 
\label{sec:exp}

The physics described above may soon be realised experimentally thanks to recent proposals for how 2D Rashba spin-orbit coupling may be added to an ultracold gas\cite{campbell, dalibard,anderson_2013, xuueda}. A Zeeman term could then be applied using an external magnetic field, while a harmonic trap can be straightforwardly added by means of additional laser beams. In this paper, we have focused on single-particle properties; these may be explored with fermionic atoms, or with species where the interaction strength can be tuned to zero by means of a Feshbach resonance. The inclusion of interactions will in general lead to other novel ground states, such as the so-called skyrmion lattice phases\cite{sinha_trapped_2011, ramachandhran_half-quantum_2012, cong-jun_unconventional_2011, hu_spin-orbit_2012, chen_2014}. 

There is also great interest in studying spin-orbit-coupled systems in photonic systems. Recently, for example, a novel spin-orbit coupling Hamiltonian was experimentally realised for polaritons in a hexagon of coupled micropillars\cite{sala}. If further advances create Rashba spin-orbit coupling, a harmonic potential could be added in the cavity arrays of experiments like Ref.~\onlinecite{sala, jacqmin} by letting the cavity size vary spatially. Photonics experiments also provide full access to the wave function, which can be imaged in real (momentum) space using the near-field (far-field) emitted light~\cite{rmp2013, jacqmin}. 
  
We now explore how observations of relevant quantities may be used to study the momentum-space artificial magnetic fields discussed above. We focus in turn on the single minimum and ring minima regimes for the lowest energy band. 

{\it Single Minimum Regime}-- In this regime, the momentum-space magnetic field breaks the degeneracy of states with the same azimuthal angular momentum. As previously studied, the effect of the Berry curvature could be measured experimentally in the dipole mode splitting of an ultracold gas even with interactions\cite{duine, pricemodes}. Collective modes are powerful tools for probing ultracold gases, as the oscillation frequencies can be measured with high precision\cite{pethick, altmeyer}. For $0<\zeta <1$ and $\chi \lesssim 1$, the upper bound on the dipole mode splitting for a spin-1/2 particle is $\delta \omega / \omega \lesssim 50\%$\cite{pricemodes}. As we have noted above, for a higher spin the ``cyclotron" frequency scales with $F$, and the dipole splitting increases. However, to satisfy the single band approximation, $ \chi \lesssim 1/F$, and so the overall upper bound on the dipole mode splitting is not increased by higher spins.  

The momentum-space magnetic field also affects the localisation of the low energy momentum-space wave functions. As discussed above, the momentum-space Fock-Darwin states are 2D harmonic oscillator states with a modified characteristic momentum scale (\ref{eq:length2}). This characteristic scale could be probed in experiments via time-of-flight measurements which map out the momentum distribution of the atomic cloud. The scale is governed by two competing effects as $\zeta$ is tuned; firstly, the effective mass decreases, which spreads the wave function out in momentum space, and secondly, the momentum-space magnetic field increases, which localises the wave function, reducing its spread. 

For $0<\zeta <1$ and $\chi \lesssim 1/F $, the net effect of changing the spin-orbit coupling strength is always to increase the characteristic momentum scale as the effective mass dominates. However, the effect of the artificial magnetic field could be isolated by measuring the variation in the spread of the wave function as $\chi$ is varied. In the limit that $\zeta \rightarrow 1$, doubling $\chi$ would reduce the momentum-space width of a wave function by a factor of $ 1/ \sqrt{2}$. 

{\it Ring Minima Regime}--The momentum-space magnetic flux can be measured in the dependence of the single-particle energy spectrum on $m$. In  general, the energy splitting of states with $+m$ and $-m$ is maximal when the Zeeman field is tuned to zero, then it is $\delta \omega / \omega = 2 m F {\chi }/ {\zeta} $. For a spin-1/2 particle we require that $\zeta / \chi \gtrsim 2$, and so the splitting can be up to: $\delta \omega / \omega  \lesssim 50\% \times m$. For a higher spin $F$ particle, the system is in Regime 3 discussed above for a vanishing Zeeman field and the single-band approximation requires that Eq. \ref{eq:regime3} is satisfied. As a result the maximum splitting between states with $+m$ and $-m$ is: 
\begin{eqnarray}
\frac{\delta \omega }{ \omega } \lesssim m F  \left(1 - \frac{(F-1)^2}{F^2} \right) 
\end{eqnarray} 
which can be a significant percentage of the harmonic trapping frequency for suitable values of $m$. Another key experimental signature of the momentum-space artificial magnetic flux is the characteristic jumps in the azimuthal angular momentum of the ground state as a function of momentum-space magnetic flux (Fig. \ref{fig:higher}). These transitions could be inferred, for example, from the variation in the ground state energy with the flux. 

As in real-space rings, the momentum-space magnetic flux induces equilibrium persistent ``currents" around the ring of minima even when the azimuthal angular momentum quantum number $m=0$.  Physically, the origin of this analogue magnetic phenomenon is from the balance of different spin components in the ground state, which carry different amounts of angular momentum. 

Persistent currents in momentum space could be investigated, for example, in the semiclassical dynamics of a wavepacket. We assume the wavepacket is prepared at an angle, $\varphi_c$, on the 1D ring in momentum space, and centred around an angular momentum, $m_c$. In the semiclassical approximation, the wavepacket moves around the momentum space ring at a rate:
\begin{eqnarray}
p_0 \dot{\varphi}_c =  \frac{\kappa}{p_0} \left(m_c - \frac{\Phi }{ \Phi_0} \right) . 
\end{eqnarray}
For a wavepacket centred around $m_c =0$, this increases as $ | \Phi / \Phi_0 | \rightarrow F$. In units of the trap frequency, the rate at which the wave packet moves around the momentum space ring of minima scales as $\frac{1}{2} \sqrt{\chi/\zeta}$ in the limit of a vanishing Zeeman field for a spin-1/2 particle. To justify our approximations, we also require that $ \zeta /\chi  \gtrsim 2 $ in the same limit, as discussed above. Then the rate would be $p_0 \dot{\varphi}_c  \lesssim \frac{1}{2\sqrt{2}} \omega$. Alternatively, persistent currents could be studied via the real-space angular momentum of the wave function or inferred from measuring the number of atoms in the different spin components. 

\section{Conclusions}

In this paper, we have shown that a single-particle with 2D spin-orbit coupling in a weak external harmonic trap can be described by an effective Hamiltonian in which the Berry curvature acts as an artificial momentum-space magnetic field. When the spin-orbit coupling strength is weak compared to the Zeeman field, the effective Hamiltonian is analogous to that of the Fock-Darwin Hamiltonian for a particle in a real space harmonic trap and uniform magnetic field. From this analogy, we have translated results more usually applied to quantum dots to describe the momentum-space properties of a spin-orbit-coupled atom. 

In the opposite limit of strong spin-orbit coupling or a weak Zeeman field, we have shown that the effective Hamiltonian is analogous to that of a particle confined to a 1D ring pierced by a real-space magnetic flux. Guided by this, we identify magnetic phenomena in momentum space, including a contribution to the energy spectrum which is (almost) periodic as a function of flux and persistent ``currents" around the momentum-space ring of energy minima. We have also extended our approach to higher spin systems, and discussed relevant experimental considerations for observing these effects. 

\acknowledgments{We are grateful to L. Santos for useful discussions. This work was funded by the Autonomous Province of Trento, partially through the Call ``Grandi Progetti 2012", Project ``On silicon chip quantum optics for quantum computing and secure communications - SiQuro", by ERC through the QGBE grant; and by EPSRC Grant EP/K030094/1.}

\appendix
\section{Energy Dispersion and Berry Curvature in Higher Spin Systems} \label{app:energy}

We consider a higher-spin system with a single-particle Hamiltonian without a harmonic trap, of the form: 
\begin{align}
	\mathcal{H}_0
	&=
	\frac{p_x^2+p_y^2}{2m} +
	\frac{1}{F} \left( a_x(\mathbf{p})F_x + a_y(\mathbf{p})F_y + a_z(\mathbf{p})F_z \right) \nonumber \\
	&\equiv
	\frac{p^2}{2m} +
	\frac{1}{F} \mathbf{a}(\mathbf{p}) \cdot \mathbf{F}, \label{ham}
\end{align}
where $F_i$ is the $i$-th component spin matrix with total spin $F$. We note that this is not the most general form of a higher-spin Hamiltonian, but can naturally include both the Zeeman field and Rashba spin-orbit coupling in the vector $\mathbf{a}(\mathbf{p}) $. The factor of $1/F$ in the second term of the Hamiltonian is inserted for convenience without loss of generality. 

Choosing the quantization axis of the spin along $\mathbf{a}(\mathbf{p})$ (called here the $z^\prime$-axis), the Hamiltonian is:
\begin{align}
	\mathcal{H}_0
	=
	\frac{p^2}{2m} + \frac{1}{F}|\mathbf{a}(\mathbf{p})|F_{z^\prime}.
\end{align}
where the matrix $F_{z^\prime}$ has the following form:
\begin{align}
	F_{z^\prime}
	=
	\begin{pmatrix}
	F & 0 & 0 & \cdots & 0 \\
	0 & F-1 & 0 & \cdots & 0 \\
	0 & 0 & F-2 & \cdots & 0 \\
	\vdots & \vdots & \vdots & \ddots & 0 \\
	0 & 0 & 0 & \cdots & -F 
	\end{pmatrix}.
\end{align} 
The energy of the $\alpha$-th band is:
\begin{align}
	E_\alpha (\mathbf{p}) = \frac{p^2}{2m} + \frac{m_\alpha}{F}|\mathbf{a}(\mathbf{p})|, \label{energy}
\end{align}
where $ (m_1, m_2, \cdots,  m_{2F+1})= (	-F, -F+1, \cdots,  F)$. For the specific case of Rashba spin-orbit coupling with a Zeeman field, the vector $
	(a_x, a_y, a_z) = (-\lambda p_y, \lambda p_x, -\Delta)$ and the energy of the $\alpha$-th band is
\begin{align}
	E_\alpha (\mathbf{p}) = \frac{p^2}{2m} + \frac{m_\alpha}{F}\sqrt{\lambda^2 (p^2) + \Delta^2}.
\end{align}
as is used in the main text. 

To calculate the Berry curvature, we express the $\alpha$-th band eigenstate of the general higher spin system in terms of a series of rotations from the spin-quantization axis, $z^\prime$: 
\begin{align}
	|u_\alpha \rangle = e^{-i\phi F_z} e^{-i\theta F_y}|m_\alpha \rangle,
\end{align}
where $F_z |m_\alpha\rangle = m_\alpha |m_\alpha\rangle$ and the angles, $\theta$ and $\phi$ are the spherical polar coordinate angles of $\mathbf{a}(\mathbf{p})$. The derivatives of $|u_\alpha \rangle$ can be expressed as: 
\begin{align}
	\left| \frac{\partial u_\alpha}{\partial p_x}\right\rangle
	&=
	-i e^{-i\phi F_z}
	\left[
	\frac{\partial \phi}{\partial p_x}F_z  
	+\frac{\partial \theta}{\partial p_x} F_y
	\right]e^{-i\theta F_y} |m_\alpha\rangle,
	\notag \\
	\left| \frac{\partial u_\alpha}{\partial p_y}\right\rangle
	&=
	-i e^{-i\phi F_z}
	\left[
	\frac{\partial \phi}{\partial p_y}F_z  
	+\frac{\partial \theta}{\partial p_y} F_y
	\right]e^{-i\theta F_y} |m_\alpha\rangle.
\end{align}
Hence, the Berry curvature for the $\alpha$-th band is: 
\begin{align}
	\Omega_\alpha (\mathbf{p})
	&=
	i
	\left[
	\left\langle \frac{\partial u_\alpha}{\partial p_x}
	\right|\left. \frac{\partial u_\alpha}{\partial p_y}\right\rangle
	-
	\left\langle \frac{\partial u_\alpha}{\partial p_y}
	\right|\left. \frac{\partial u_\alpha}{\partial p_x}\right\rangle
	\right]
	\notag \\
	&=
	-
	\left(
	\frac{\partial \theta}{\partial p_x}\frac{\partial \phi}{\partial p_y}
	-\frac{\partial \phi}{\partial p_x}\frac{\partial \theta}{\partial p_y} 
	\right)
	\langle m_\alpha| e^{i\theta F_y} F_x e^{-i\theta F_y} |m_\alpha\rangle. \notag
\end{align}
Using the Baker-Campbell-Hausdorff lemma
$e^X Y e^{-X} = Y + [X,Y] + \frac{1}{2!}[X,[X,Y]] + \frac{1}{3!}[X,[X,[X,Y]]] + \cdots$,
and the spin algebra $[F_i, F_j] = i\epsilon_{ijk}F_k$, one can show that: 
\begin{align}
	e^{i\theta F_y} F_x e^{-i\theta F_y}
	=
	F_x \cos \theta + F_z \sin \theta.
\end{align}
Since $F_x = (F_+ + F_-)/2$, where $F_\pm$ are ladder operators, $\langle m_\alpha | F_x |m_\alpha\rangle = 0$.
Therefore: 
\begin{align}
	\Omega_\alpha (\mathbf{p})
	&=
	-
	\left(
	\frac{\partial \theta}{\partial p_x}\frac{\partial \phi}{\partial p_y}
	-\frac{\partial \phi}{\partial p_x}\frac{\partial \theta}{\partial p_y} 
	\right)
	\langle m_\alpha| F_z |m_\alpha\rangle \sin \theta
	\notag \\
	&=
	-
	\left(
	\frac{\partial \theta}{\partial p_x}\frac{\partial \phi}{\partial p_y}
	-\frac{\partial \phi}{\partial p_x}\frac{\partial \theta}{\partial p_y} 
	\right)
	m_\alpha
	\sin \theta \label{curvature}. 
\end{align}
For the specific case of a particle with Rashba spin-orbit coupling and a Zeeman field, the rotation angles can be calculated from $a({\bf p})$  as:
\begin{align}
	\sin \phi &= \frac{p_x}{\sqrt{p_x^2 + p_y^2}}, &
	\sin \theta &= \frac{\lambda \sqrt{p_x^2 + p_y^2}}{\sqrt{\lambda^2 (p_x^2+p_y^2) + \Delta^2}}. 
\end{align}
Hence the Berry curvature of the $\alpha$-th band is\cite{berry}: 
\begin{align}
	\Omega_\alpha (\mathbf{p})
	=
	\frac{\lambda^2 m_\alpha \Delta}{\left( \lambda^2 (p_x^2+p_y^2) + \Delta^2 \right)^{3/2}},
\end{align}
as stated in the main text.

\end{document}